\documentclass[sigconf,nonacm]{acmart} % arxiv
\usepackage{array}
\usepackage{fvextra}
\usepackage{booktabs}
\usepackage{tabularx}
\usepackage{enumitem}
\usepackage{listings}
\usepackage{appendix}
\usepackage{tikz}

\DefineVerbatimEnvironment{promptblock}{Verbatim}{
  breaklines=true,
  breakanywhere=true,
  fontsize=\footnotesize,
  baselinestretch=1,
  xleftmargin=0.5em,
  frame=single,
  framesep=3mm
}

\usepackage{graphicx}
\newcommand{\TagFile}[1]{images/tags/#1.pdf}
\newcommand{\GreyTagFile}[1]{images/tags/grey_#1.pdf}
\newcommand{\GreenTagFile}[1]{images/tags/green_#1.pdf}
\newcommand{\CircleTagFile}[1]{images/tags/circle_#1.pdf}
\newcommand{\pdfnum}[1]{\raisebox{-0.3ex}{\includegraphics[height=2.2ex]{\TagFile{#1}}}}
\newcommand{\greynum}[1]{\raisebox{-0.5ex}{\includegraphics[height=2.4ex]{\GreyTagFile{#1}}}}
\newcommand{\greennum}[1]{\raisebox{-0.5ex}{\includegraphics[height=2.4ex]{\GreenTagFile{#1}}}}
\newcommand{\pdfalpha}[1]{\raisebox{-0.5ex}{\includegraphics[height=2.4ex]{\CircleTagFile{#1}}}}

\newcolumntype{Y}{>{\centering\arraybackslash}X}
\AtBeginDocument{%
  }

\setcopyright{none}
\copyrightyear{2026}
\acmYear{2026}
\acmDOI{}
\acmConference[UIST '26]
{The 39th Annual ACM Symposium on User Interface Software and Technology}
{November 2--5, 2026}
{Detroit, MI, USA}
\acmISBN{}

\newcommand{\oursystem}{\textit{OrchestrXR}}

\begin{document}

\title{\oursystem{}: A Multi-Agent System for Idea-to-Prototype \\ XR Study Authoring}

% \affiliation{%

\author{Shuqi Liao}
\affiliation{%
\institution{Purdue University}
\city{West Lafayette}
\state{Indiana}
\country{USA}}

\author{Chenfei Zhu}
\affiliation{%
\institution{Purdue University}
\city{West Lafayette}
\state{Indiana}
\country{USA}}

\author{Karthik Ramani}
\affiliation{%
\institution{Purdue University}
\city{West Lafayette}
\state{Indiana}
\country{USA}}

\author{Voicu Popescu}
\affiliation{%
\institution{Purdue University}
\city{West Lafayette}
\state{Indiana}
\country{USA}}

% \affiliation{%

% \affiliation{%

% \affiliation{%


\begin{abstract}
  Extended Reality (XR) has become an important interaction paradigm in Human-Computer Interaction (HCI). XR studies are used to investigate interaction, perception, and user behavior in immersive environments, and typically involve experimental tasks, 3D scenes, and interactive logic. However, turning an initial XR study idea into a runnable prototype remains fragmented across study design, scene construction, and interaction implementation. We present \oursystem{}, a multi-agent human--AI workflow for early-stage idea-to-prototype XR study authoring. Rather than treating XR study creation as one-shot generation, \oursystem{} supports a controllable workflow across study design, scene generation, and interaction generation through structured schemas, multi-agent orchestration, and interactive human-agent interfaces, producing a Unity-based prototype from a researcher’s idea. A user study with 12 XR researchers suggests that \oursystem{} provides effective support for early-stage XR study authoring with strong intent preservation across stages.
\end{abstract}

\begin{CCSXML}
<ccs2012>
 <concept>
  <concept_id>10003120.10003121.10003124.10010865</concept_id>
  <concept_desc>Human-centered computing~Virtual reality</concept_desc>
  <concept_significance>500</concept_significance>
 </concept>
 <concept>
  <concept_id>10003120.10003121.10003125.10010875</concept_id>
  <concept_desc>Human-centered computing~Interactive systems and tools</concept_desc>
  <concept_significance>300</concept_significance>
 </concept>
 <concept>
  <concept_id>10010147.10010257.10010293.10010294</concept_id>
  <concept_desc>Computing methodologies~Intelligent agents</concept_desc>
  <concept_significance>300</concept_significance>
 </concept>
</ccs2012>
\end{CCSXML}

\ccsdesc[500]{Human-centered computing~Interactive systems and tools}
\ccsdesc[300]{Human-centered computing~Virtual reality}
\ccsdesc[300]{Computing methodologies~Intelligent agents}

\keywords{multi-agent systems, large language models, authoring tools, extended reality, human-AI collaboration}

\begin{teaserfigure}
  \includegraphics[width=\textwidth]{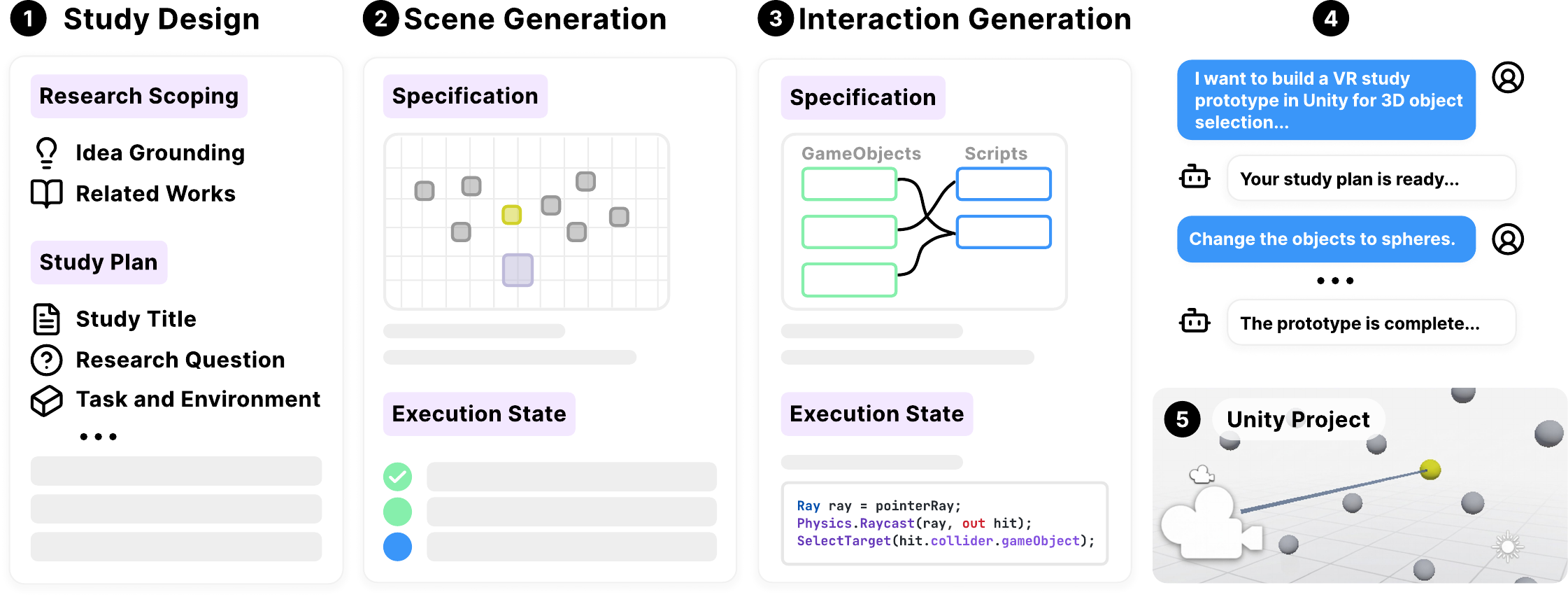}
  \caption{\oursystem{} structures XR study authoring into three connected agent stages: \pdfnum{1} Study Design (SD), \pdfnum{2} Scene Generation (SG), and \pdfnum{3} Interaction Generation (IG). Users interact with the system through a shared \pdfnum{4} chat interface, while \pdfnum{5} Unity serves as the execution environment. Starting from an XR study idea, the system progressively translates research intent into structured study, scene, and interaction specifications, and ultimately an executable XR prototype in Unity.}
  \Description{xxx}
  \label{fig:teaser}
\end{teaserfigure}


\maketitle

\section{Introduction}

% 1. talk about current xr research workflow exsit fragmentation. 
% 2. The AI opportunity and its Gap to solve this fragmentation
% 3. XResearcher: how to bridge the gap from three levels: design from representation (data), agent orchestration (backend), and user interfaces (frontend)
% 4. Summary of the Evaluation (User Study)
% 5. Contributions

Extended Reality (XR) has become an important interaction paradigm in Human-Computer Interaction (HCI)~\cite{milgram1994taxonomy, speicher2019mixed}. XR studies are commonly used to investigate interaction, perception, and user behavior in immersive environments, and typically involve experimental tasks, 3D scenes, and interactive logic~\cite{pan2018why, ratcliffe2021extended}. However, turning an initial XR study idea into a runnable prototype remains difficult~\cite{ashtari2020creating, krauss2021current}. Authoring such prototypes requires researchers to move across disconnected layers of work, from specifying study goals and procedures, to constructing 3D scenes, to implementing interactive behavior in engines such as Unity~\cite{unity_engine} or Unreal Engine~\cite{unreal_engine}. Because this process is also inherently iterative, pilot findings often force revisions across all three layers. As a result, early-stage XR prototyping remains fragmented, leaving a substantial gap between research ideas and executable artifacts.

Recent advances in Large Language Models (LLMs) have created new opportunities for complex authoring workflows, especially through multi-agent systems that can coordinate specialized roles and intermediate artifacts. Prior work has begun to explore LLMs and multi-agent systems for automating scientific workflows~\cite{ghafarollahi2024protagents, schmidgall2025agent, yamada2025ai}. However, most existing approaches primarily operate on text or 2D artifacts~\cite{zhou2023webarena, yang2024swe, deng2023mind2web, qin2025ui}. XR study authoring, by contrast, requires researchers to bridge multiple representations from abstract study design to spatial scene specification and executable interaction logic. This gap suggests an opportunity to leverage LLM-based agentic support for XR study authoring, helping accelerate early-stage XR research by reducing low-level implementation burden and enabling researchers to focus on higher-level idea creation.

To ground XR study authoring support in real practice, we conducted a formative study with six XR researchers. We found that XR study development is highly iterative and fragmented, requiring researchers to formalize underspecified ideas, translate study intent across representations, and coordinate work across multiple tools while preserving experimental control. These findings led to three design strategies for AI-assisted XR study authoring: \textit{structured schemas} for stage-specific authoring artifacts, \textit{multi-agent orchestration} across connected workflow stages, and \textit{interactive human-agent interfaces} that support inspection and control throughout the authoring process.

Guided by these three strategies, we present \oursystem{}, a multi-agent authoring system for early-stage XR study prototyping in Unity. To the best of our knowledge, \oursystem{} is the first framework to integrate LLM-based multi-agent support into a unified workflow for XR study authoring, spanning study design, scene generation, and interaction generation. \oursystem{} enables researchers to progressively transform study ideas into inspectable design, scene, and interaction artifacts under user control.

We evaluate \oursystem{} through a user study with 12 XR researchers. Participants each completed two assigned task cases, balanced across four representative XR study authoring scenarios, followed by an open-ended exploration session using their own XR study ideas. Results suggest that \oursystem{} effectively helps transform abstract XR study ideas into inspectable prototype artifacts, preserve core study intent across authoring stages, and supports a coherent workflow from idea to runnable prototype.

In summary, our paper makes the following contributions: \textbf{(1)} \oursystem{}, a multi-agent framework that unifies study design, scene generation, and interaction generation for idea-to-prototype XR study authoring through structured intermediate artifacts and a controllable workflow. \textbf{(2)} An empirical evaluation with 12 XR researchers across four XR study cases, demonstrating the effectiveness of \oursystem{}, its positive reception among XR researchers, and implications for future XR authoring tools.
\section{Related Works}

\subsection{XR Study Support}

Several prior systems support replay, inspection, and post-hoc analysis of XR study data~\cite{hubenschmid2022relive,javerliat2024plume,nebeling2020mrat}. For example, ReLive~\cite{hubenschmid2022relive} combines in-situ VR replay with ex-situ desktop analytics for synchronized inspection of MR sessions and multimodal data. These systems mainly help researchers analyze and interpret XR studies after data collection, whereas \oursystem{} focuses on an earlier phase by helping researchers externalize an initial study idea into an inspectable and executable XR prototype.

Other work supports the specification and execution of XR experiments through reusable experiment frameworks~\cite{picard2024xrmuse, ehret2024studyframework,brookes2020studying,bebko2020bmltux}. In particular, StudyFramework~\cite{ehret2024studyframework} simplifies factorial-design VR studies in Unreal through condition generation, counterbalancing, logging, and experimenter monitoring, while prior Unity-based toolkits similarly provide reusable support for experiment structure, settings, and data collection~\cite{brookes2020studying,bebko2020bmltux}. In contrast, these systems primarily assume that core study structure has already been defined, whereas \oursystem{} targets the earlier authoring step of progressively deriving study design, scene, and interaction specifications from a high-level XR research idea.

Some systems extend XR studies with additional runtime capabilities, such as remote supervision, distributed execution, monitoring, and in-situ experiment support~\cite{lee2022remotelab,steed2022ubiq,stefanidi2022argus}. For example, RemoteLab~\cite{lee2022remotelab} supports supervised remote VR experiments with synchronized multi-site control, monitoring, and recording. Together, these systems broaden how XR studies can be run in practice, while \oursystem{} contributes complementary support at the front end of the workflow by helping researchers author the study prototype itself before deployment.

\subsection{AI-based Research Tool}

A related line of work explores AI support for multi-stage research workflows~\cite{lu2024ai,schmidgall2025agent}. For example, the AI Scientist~\cite{lu2024ai} combines idea generation, code-based experimentation, figure creation, paper drafting, and automated review in a closed-loop workflow. Subsequent work extends this paradigm through stronger search, verification, and domain specialization~\cite{yamada2025ai,ifargan2025autonomous,li2024mlr,gottweis2025towards}. These systems share our interest in structured AI support across multiple research stages, but they primarily target general scientific research rather than XR study authoring and executable prototype creation.

Another line of work focuses more specifically on literature-centered research support. PaperQA~\cite{lala2023paperqa} and PaSa~\cite{he2025pasa} frame literature interaction primarily as retrieval-grounded question answering and scholarly search, while other systems support literature sensemaking, organization, and synthesis~\cite{kang2023synergi,wang2024scidasynth,ma2025garden,zheng2024disciplink,skarlinski2024language}. These systems show how LLM-based tools can assist upstream research understanding and synthesis, whereas \oursystem{} focuses on transforming research intent into concrete study specifications and prototype artifacts for XR workflows.

\subsection{Multi-Agent System}

Recent LLM-based multi-agent systems provide reusable coordination mechanisms such as role specialization, delegation, hierarchical planning, and structured inter-agent communication~\cite{wu2024autogen,hong2023metagpt,chen2023agentverse,fourney2411magentic,gao2024agentscope}. Such coordination patterns have also begun to appear in LLM-based Agentic XR- and AR-oriented authoring systems~\cite{de2024llmr,lee2025imaginatear,zhu2025agentar,carcangiu2025tellxr,mereu2024empowering,li2025xr,du2026vibe}, and inform the staged design of \oursystem{}.

Multi-agent systems have also been applied to research support, reasoning, and software engineering. Prior work includes agentic research assistants~\cite{lu2024ai,schmidgall2025agent,hou2025paperdebugger}, debate- or consensus-based reasoning systems~\cite{du2024improving,liang2024encouraging,li2024improving,estornell2024multi}, and role-based software development workflows such as ChatDev~\cite{qian2024chatdev}, CodeR~\cite{chen2024coder}, MAGIS~\cite{tao2024magis}, and UniDebugger~\cite{lee2025unidebugger}. Related workflow-oriented systems also treat multi-agent coordination as one strategy within broader human-AI support pipelines~\cite{kang2023synergi,xie2024waitgpt,overney2024sensemate}. Taken together, these works motivate multi-agent orchestration as a practical strategy for decomposing complex tasks; \oursystem{} extends this idea to early-stage XR study authoring through inspectable, human-steerable stages that progressively lead to a Unity-based prototype.

\section{Formative Study}
To understand the challenges XR researchers face in designing XR studies and to identify design implications for assistive authoring tools, we conducted a formative study using semi-structured interviews.

\subsection{Participants and Procedure}
We recruited 6 XR researchers from research groups at a public research university, including Ph.D. students and faculty members with experience independently designing and conducting XR studies. Their research interests spanned education, virtual agents, rendering, and collaborative work. Detailed demographics are provided in Appendix~\ref{app:formative_demo}.

Each remote interview lasted about 30 minutes. Before the interview, participants completed a short survey on demographics, research areas, and typical XR study goals. We then followed a semi-structured protocol covering their typical XR study authoring workflow from initial ideas to runnable implementations, the challenges they encountered, and the strategies or tools they used to address them. Participants were encouraged to discuss recent projects and concrete examples of breakdowns during design and implementation. All interviews were audio-recorded with consent and transcribed for analysis.

\subsection{Findings}
\label{sec:findings}

Two researchers analyzed the transcripts through iterative coding and discussion, grouping recurring challenges into higher-level themes by consensus. This process yielded five key findings about common challenges in XR study authoring \textbf{(F1--F5)}.

\paragraph{\textbf{F1: Early-stage XR study designs are underspecified for execution.}}
Participants described initial research ideas as often remaining at a high level (e.g., a general question or comparison) and lacking the concrete specifications needed for implementation. As a result, they must iteratively refine variables, conditions, tasks, and procedures before a runnable XR study can be developed. As P1 noted, ``\emph{you have to respect existing research design and analysis frameworks... you cannot just invent your own trick}.''

\paragraph{\textbf{F2: XR study design requires translating conceptual intent into spatial and executable representations.}}
Participants highlighted the difficulty of mapping abstract experimental intent (e.g., variables and conditions) into concrete XR environments and executable logic. This translation spans study constructs, spatial scenes, interactions, and procedural implementation, and often requires early feasibility assessment. As P4 noted, after reviewing prior work, they consider whether an idea is ``\emph{doable from the level of implementation},'' while P5 emphasized that some projects require ``\emph{the algorithm... that simulates this behavior}.''

\paragraph{\textbf{F3: XR study authoring is fragmented across tools, and current AI support is often local rather than global.}}
Participants described XR study authoring as distributed across disconnected artifacts and platforms, including literature notes, design documents, game engines, and code, which increases coordination overhead and context switching. They also noted that current AI support is often effective only for localized sub-tasks and struggles to maintain project-level coherence. As P4 explained, models can help generate code only when enough detailed context is provided, but otherwise ``\emph{cannot give you the exact implementation},'' and more broadly ``\emph{cannot comprehend a comprehensive idea about your project... and have trouble to connect your pieces together}.''

\paragraph{\textbf{F4: Study design and implementation are tightly coupled and evolve iteratively.}}
Participants consistently described XR study authoring as iterative rather than linear. Early prototypes and implementations are used to validate feasibility and refine study design, and unexpected outcomes can lead to changes in both the implementation and the research itself. As P2 noted, ``\emph{only with a real implementation we can know that the idea actually works or not},'' while P5 described revising the system and ``\emph{maybe slightly change even the research}.''

\paragraph{\textbf{F5: Maintaining controlled and reusable experimental setups is critical but challenging.}}
Participants emphasized the importance of keeping experimental setups stable and reusing existing environments to isolate variables and support valid comparisons. Rather than rebuilding scenes, they preferred to change only the factors under study, since broader modifications could introduce confounds. As P5 explained, ``\emph{we try to keep things stable, and only manipulate the things we want to explore},'' because changing the environment ``\emph{immediately impacts the users}.''

\begin{figure*}[t!]
    \centering
    \includegraphics[width=0.7\textwidth]{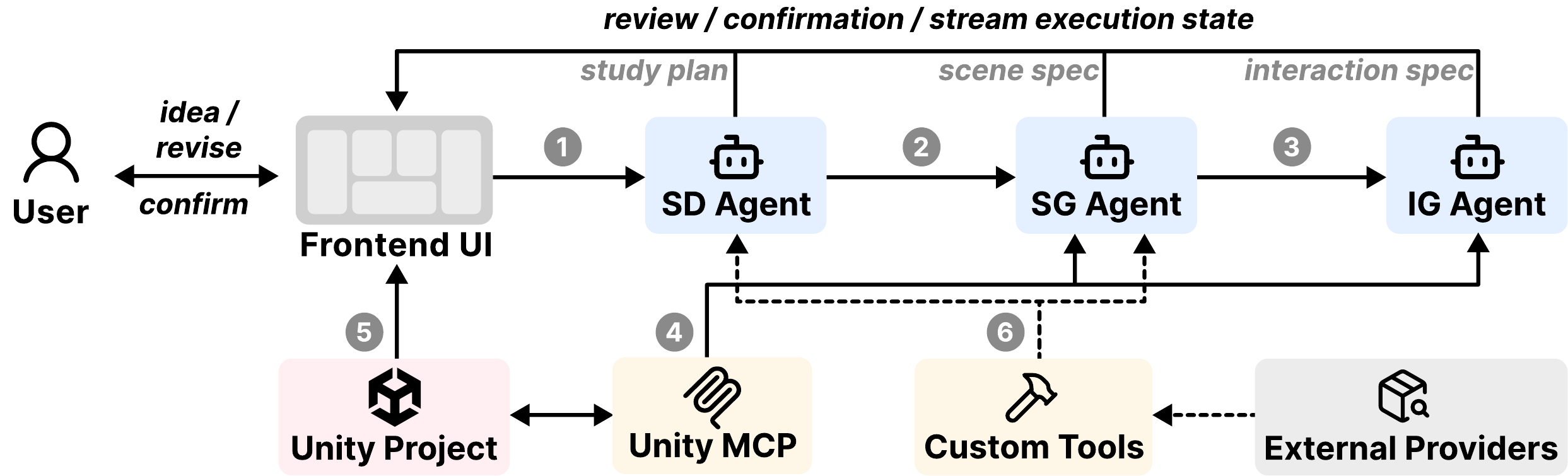}
    \caption{\oursystem{} system workflow. The system is organized as a stage-based authoring process in which the frontend mediates user interaction with three connected backend agents: Study Design (SD), Scene Generation (SG), and Interaction Generation (IG). Intermediate artifacts are progressively passed across stages, with user revision and confirmation interleaved throughout. Gray numbered circles indicate the main data flow described in Sec.~\ref{sec:system_overview}. Arrows denote data-flow direction; communication with Unity is bidirectional, while other tool and provider connections are one-way.}
    \label{fig:system_workflow}
\end{figure*}

\subsection{Design Considerations}
\label{sec:design_considerations}

Based on these findings, we derive the following design considerations for tools that support XR study authoring.

\paragraph{\textbf{DC1. Establish a unified XR study representation as the single source of truth.}}
To support underspecified early-stage designs (F1) and the translation from conceptual intent to spatial and executable forms (F2), assistive tools should provide a unified representation of an XR study that captures core elements such as research intent, variables, conditions, tasks, and procedures in a structured form. This representation should serve as a stable single source of truth that can be progressively refined and used to drive downstream environment and implementation artifacts.

\paragraph{\textbf{DC2. Provide an integrated authoring system that bridges ideation, XR environment design, and implementation.}}
Because XR study authoring is fragmented across tools and representations (F3), tools should support the end-to-end workflow from research reasoning to runnable study setup within a unified authoring environment. Rather than forcing users to move across disconnected documents, engines, and codebases, the system should coordinate authoring across stages in one interface.

\paragraph{\textbf{DC3. Make cross-layer mappings explicit to support coherent AI assistance and user control.}}
Because translating study intent into XR realizations involves different representations (F2) and current AI assistance is often local rather than global (F3), tools should make the mapping between study elements and their realizations explicit and inspectable. This can help maintain project-level coherence, reduce the need to repeatedly restate context, and let users review and revise intermediate specifications before committing to generated environments or code.

\paragraph{\textbf{DC4. Support iterative refinement while preserving experimental control through reuse and stable baselines.}}
Since study design and implementation co-evolve iteratively (F4) and researchers prioritize stable, reusable setups to isolate variables (F5), tools should support iteration without disruptive rework. This includes enabling controlled modifications to specific factors while keeping other aspects stable, and supporting reuse of environments and study components to preserve experimental validity across iterations.
\begin{figure*}
    \centering
    \includegraphics[width=1\textwidth]{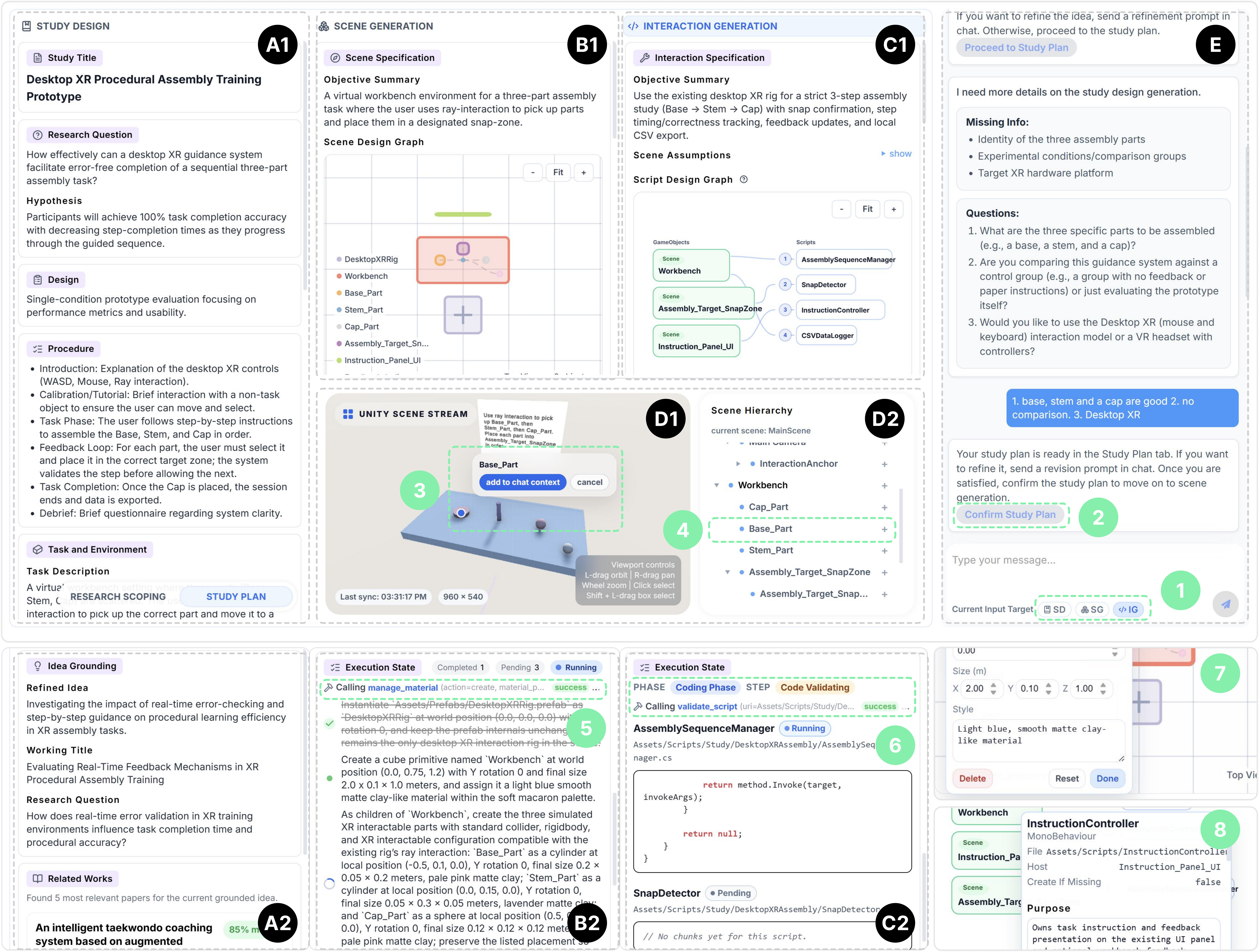}
    \caption{\oursystem{} frontend interface. A unified web interface supports the full authoring workflow across five areas: Study Design (A), Scene Generation (B), Interaction Generation (C), Unity Scene Stream (D), and a human–agent chat window (E). Users chat with agents by selecting the target context (\greennum{1}) and confirming stage progression (\greennum{2}).}
    \label{fig:interface_overview}
\end{figure*}

\section{\oursystem{} System Design}

\subsection{Design Strategies}
Grounded in the design considerations derived from our formative study (Section~\ref{sec:design_considerations}), \oursystem{} is organized around three complementary design strategies: \textit{Structured Schemas}, \textit{Multi-Agent Orchestration}, and \textit{Interactive Human-Agent Interfaces}. Together, these strategies support a unified authoring system that links early-stage XR research ideation to executable prototype generation while preserving user oversight throughout the process.

\textbf{\textit{Structured Schemas.}} To establish a unified XR study representation and make cross-layer mappings explicit (\textbf{\textit{DC1}}, \textbf{\textit{DC3}}), \oursystem{} uses structured schemas as stage-specific representations throughout the authoring process. Instead of relying on loosely connected prompts and outputs, the system maintains explicit intermediate representations for study, scene, and interaction specifications across stages. These schemas carry key study elements in a stable and inspectable form, reducing ambiguity in early-stage ideas while preserving coherence across the workflow. Structured schemas therefore serve as the representational backbone of the system, making study intent legible, transferable, and revisable across heterogeneous authoring layers.

\textbf{\textit{Multi-Agent Orchestration.}} To bridge ideation, XR environment design, and implementation within a coordinated workflow (\textbf{\textit{DC2}}), \oursystem{} decomposes authoring into three connected stages---study design, scene generation, and interaction generation---each supported by a specialized agent. Rather than providing only local assistance on isolated subtasks, the system propagates context and intermediate artifacts across stages so that higher-level research intent can inform downstream scene and interaction decisions. This strategy reduces fragmentation by connecting conceptual, spatial, and executable representations within one workflow, while supporting iterative refinement as outputs from one stage can be revised and carried forward to the next.

\textbf{\textit{Interactive Human-Agent Interfaces.}} To support inspectable AI assistance and iterative refinement under user control (\textbf{\textit{DC3}}, \textbf{\textit{DC4}}), \oursystem{} exposes intermediate artifacts and generation progress through a shared authoring interface. Instead of treating prototype generation as one-shot automation from a high-level prompt, the system supports multi-round human-agent collaboration in which users can review, revise, and confirm intermediate outputs before proceeding. This interaction model helps users maintain control over study intent, assess feasibility as the prototype evolves, and preserve stable baselines while selectively modifying specific factors. By making intermediate representations visible and actionable, the interface turns authoring into a controllable workflow rather than a black-box generation pipeline.

\subsection{System Overview}
\label{sec:system_overview}

Figure~\ref{fig:system_workflow} summarizes the overall workflow of \oursystem{}. The system is organized as a frontend-mediated authoring workflow that connects three backend agents: Study Design (SD), Scene Generation (SG), and Interaction Generation (IG). Rather than treating XR study prototyping as one-shot generation, \oursystem{} structures it as a progressive process in which intermediate artifacts are incrementally produced, inspected, and passed across agents under user supervision.

The workflow begins when the user provides an initial XR study idea through the frontend interface (Fig.~\ref{fig:interface_overview}, \pdfalpha{E}). Based on this input (Fig.~\ref{fig:system_workflow}, \greynum{1}), the SD Agent produces a structured study plan (Fig.~\ref{fig:interface_overview}, \pdfalpha{A1}) that grounds the user's research intent into an explicit XR study representation. This study plan, together with the original user idea, forms the high-level study objective (Fig.~\ref{fig:system_workflow}, \greynum{2}) and is passed to the SG Agent, which generates a scene specification (Fig.~\ref{fig:interface_overview}, \pdfalpha{B1}) describing the target XR environment. After the scene specification is confirmed by the user, the SG Agent incrementally executes the scene construction process in the Unity project, while streaming the execution state to the frontend (Fig.~\ref{fig:interface_overview}, \pdfalpha{B2}). Once the user is satisfied with the generated scene, the workflow proceeds to the IG Agent. The IG Agent takes the study design, confirmed scene specification, and live Unity scene context as input (Fig.~\ref{fig:system_workflow}, \greynum{3}) and produces an interaction specification (Fig.~\ref{fig:interface_overview}, \pdfalpha{C1}) that defines the executable interaction logic and assembly plan for the prototype. After user confirmation, the IG Agent starts implementing the interaction logic in Unity while streaming the execution state to the frontend (Fig.~\ref{fig:interface_overview}, \pdfalpha{C2}). Across these stages, outputs from one stage become structured inputs to the next, preserving contextual continuity from high-level research intent to scene and interaction realization while keeping users in the loop through inspection, revision, and confirmation.

To support executable prototype realization, the agents connect bidirectionally to the Unity project through Unity MCP~\cite{mcp_spec}, which provides tools and resources (Fig.~\ref{fig:system_workflow}, \greynum{4}) for observing and manipulating the Unity scene during scene and interaction generation. Unity scene views and hierarchy states (Fig.~\ref{fig:system_workflow}, \greynum{5}) are streamed back to the frontend (Fig.~\ref{fig:interface_overview}, \pdfalpha{D1}, \pdfalpha{D2}), allowing users to inspect and steer the generation process without switching between Unity and the web interface. In addition, customized external tools provide auxiliary support for intermediate generation, including literature retrieval for the SD Agent and asset search and download for the SG Agent (Fig.~\ref{fig:system_workflow}, \greynum{6}). Together, these components form an integrated workflow that links user-driven XR study authoring with Unity-based prototype realization.

To realize the design strategy of an \textit{interactive human-agent interface}, we design the frontend around three mechanisms. First, key intermediate representations are directly editable: users can revise the scene design graph during scene specification generation and the script design graph during interaction generation (Fig.~\ref{fig:interface_overview}, \greennum{7}, \greennum{8}). Second, we provide live Unity scene and hierarchy streams for situated grounding; users can click scene objects in the scene stream (Fig.~\ref{fig:interface_overview}, \greennum{3}) or use the add button in the hierarchy stream (Fig.~\ref{fig:interface_overview}, \greennum{4}) to insert object names directly into the prompt. Third, we expose agent progress and internal state during generation: scene generation shows plan-step status together with tool calling, reasoning, and replanning (Fig.~\ref{fig:interface_overview}, \greennum{5}), while interaction generation shows script visualization, phase status, tool calls, and reasoning (Fig.~\ref{fig:interface_overview}, \greennum{6}). Together, these mechanisms enable users to inspect intermediate artifacts, ground follow-up instructions in the current scene state, and steer agent behavior throughout the authoring process.

% \begin{figure}[t!]
%     \centering
%     \includegraphics[width=0.75\linewidth]{images/sd_agent.pdf}
%     \caption{Study Design Agent workflow. Blue nodes denote single-LLM reasoning stages, and dotted arrows denote human-in-the-loop revision paths.}
%     \label{fig:sd_agent}
% \end{figure}

% \begin{figure}[t!]
%     \centering
%     \includegraphics[width=1\linewidth]{images/sg_agent.pdf}
%     \caption{Caption}
%     \label{fig:sg_agent}
% \end{figure}

% \begin{figure}[t!]
%     \centering
%     \includegraphics[width=1\linewidth]{images/ig_agent.pdf}
%     \caption{Interaction Agent workflow.}
%     \label{fig:ig_agent}
% \end{figure}

\begin{figure}[t!]
    \centering
    \includegraphics[width=1\linewidth]{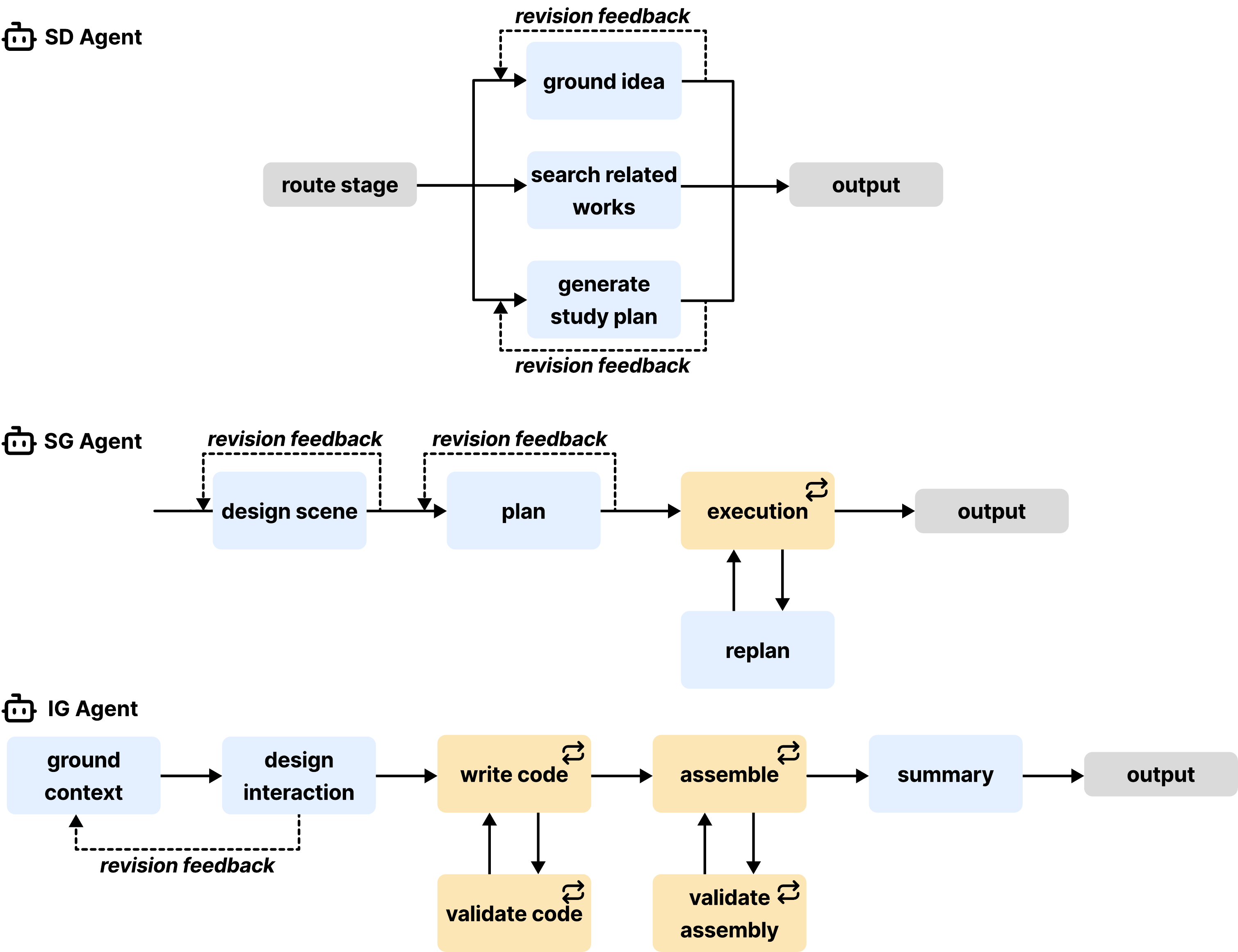}
    \caption{Workflows of the three agents: SD Agent (top), SG Agent (middle), and IG Agent (bottom). Blue nodes denote single-LLM reasoning stages, yellow nodes denote ReAct-style agent stages with attached tools, and dotted arrows indicate human-in-the-loop revision paths.}
    \label{fig:three_agents}
\end{figure}

\begin{figure*}[t!]
    \centering
    \includegraphics[width=0.7\linewidth]{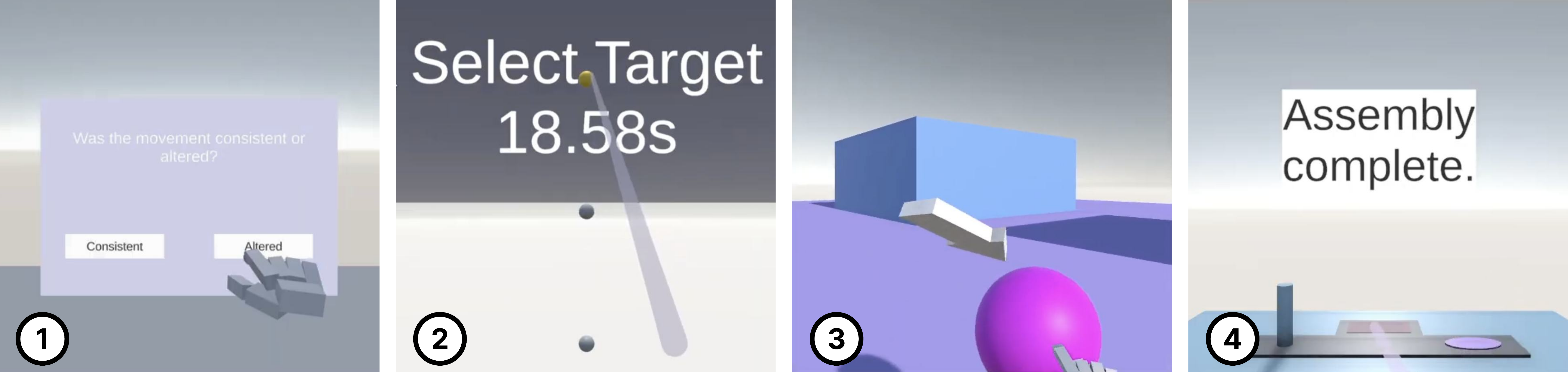}
    \vspace{-2pt}
    \caption{Results from the four case studies. The figure shows one prototype result from each case: (1) hand redirection threshold detection, (2) 3D object selection, (3) spatial search with navigation aid, and (4) procedural assembly training.}
    \label{fig:four_cases}
    \vspace{-4pt}
\end{figure*}

\subsection{Study Design Agent}

Fig.~\ref{fig:three_agents} (top) shows the workflow of the SD Agent. It takes either an initial XR study idea or a user revision request as input. The initial \textbf{route stage} determines the next operation based on the user's current intent and the current stage of study authoring. The \textbf{ground idea} node first reformulates the user's idea into a more concrete and study-oriented description, or returns clarification questions when the input is underspecified. The \textbf{search related works} node then augments the model with retrieval tools to identify relevant prior work and rank candidate papers by relevance. Based on the grounded idea and retrieved references, the \textbf{generate study plan} node synthesizes the available context into a structured study plan. The workflow also supports human-in-the-loop revision through the feedback paths, allowing users to revise intermediate outputs before the next pass; in practice, the agent is typically used progressively for idea grounding, related-work searching, and study-plan generation.

\subsection{Scene Generation Agent}

Fig.~\ref{fig:three_agents} (middle) shows the workflow of the SG Agent. The \textbf{design scene} node takes the confirmed study plan from the SD Agent and generates a structured scene specification, defining the scene objective, concrete object instances, their relative placement, and high-level layout constraints. Once the scene specification is confirmed by the user, the \textbf{plan} node produces an executable step-by-step construction plan, where each step corresponds to a concrete scene-building action. After plan approval, the \textbf{execution} node carries out these steps in Unity through grounded tool use for scene inspection and modification, while the \textbf{replan} node updates the remaining plan based on execution outcomes and the current scene state. This execution--replanning loop continues until the final scene is constructed. The \textbf{execution} node follows a ReAct-style design~\cite{yao2022react}, interleaving tool calls, observations, and step-level reasoning; to keep execution focused on the active subtask, it is conditioned only on the remaining plan steps rather than the full upstream study and scene-design context~\cite{liu2024lost}.

\subsection{Interaction Generation Agent}

Fig.~\ref{fig:three_agents} (bottom) illustrates the workflow of the IG Agent. The \textbf{ground context} node first consolidates the confirmed study plan, scene specification, and live Unity scene hierarchy into a compact context for downstream code generation and assembly. Based on this context, the \textbf{design interaction} node produces a structured interaction specification that defines the interaction objective, scene assumptions, script contracts, generation order, and assembly actions. After user approval, the \textbf{write code} node generates scripts one at a time, and the \textbf{validate code} node checks each script for contract compliance and compilation status using Unity MCP tools, looping back for revision if needed until the script set is completed or the retry limit is reached (n=3). The \textbf{assemble} node then applies the specification to the scene by creating required host objects, attaching scripts to target GameObjects, and binding scene references. The \textbf{validate assembly} node then checks the resulting scene state in Unity. If assembly validation fails, the workflow retries assembly up to three times; otherwise, it proceeds to the \textbf{summary} node, which reports generation results. The \textbf{write code}, \textbf{validate code}, \textbf{assemble}, and \textbf{validate assembly} nodes are implemented as ReAct-style agents with different tool configurations, enabling grounded generation and verification that improves reliability and supports XR study scenarios involving multiple scripts and scene objects. For reproducibility, detailed system prompts and LLM configuration settings for the nodes in each agent are provided in Appendix~\ref{app:systemprompt}.

\subsection{Patch-and-Revise Mode}
Complex Unity authoring tasks are difficult to complete reliably in a single pass. Prior work shows that contemporary tool-using agents remain brittle on long-horizon tasks: errors accumulate over extended execution chains, and execution, grounding, and recovery remain key bottlenecks~\cite{yu2026infiagent,aghzal2026why}. We therefore provide a \textit{Patch-and-Revise} mode that allows users to steer each agent after the initial generation run. In the chat box (Fig.~\ref{fig:interface_overview}, \greennum{1}), follow-up prompts can be directed to SD, SG, or IG. For SD, prompts are handled as revisions within the original study-design workflow. For SG and IG, the system distinguishes broad redesign requests from local corrections: broad requests return to the full generation graph, while local corrections invoke dedicated patch agents. The SG patch agent uses the current scene specification and a live Unity scene snapshot to make minimal scene edits without changing the overall scene goal. The IG patch agent uses the current interaction specification, generated scripts, the current scene specification and Unity scene snapshot, and any available compile error trace to minimally update scripts, scene assembly, or bindings while preserving the existing interaction architecture. Both patch agents are lightweight ReAct agents using the corresponding full-stage configurations; requests that exceed safe local modification fall back to the full revise workflow.

\section{Case Study: Reproducing XR Study Prototypes}
\label{sec:case_study}

We conducted a case study to evaluate whether \oursystem{} can support the authoring of structured XR study prototypes from high-level research ideas. Rather than directly reproducing prior systems, we drew inspiration from several representative XR studies and extracted their core study elements to design four case tasks. We then examined whether \oursystem{} could generate corresponding study designs, scene specifications, and interaction implementations from these case prompts. The full input prompts for all cases are provided in the Appendix~\ref{app:user_study_initial_prompts}.

The four cases were inspired by prior XR research spanning perception~\cite{hartfill2021analysis}, selection~\cite{dalsgaard2021modeling}, navigation~\cite{kumaran2023impact}, and procedural assembly training~\cite{carlson2015virtual}. Our goal was not exact replication, but reconstruction of the core study structure---including task flow, interaction paradigm, and measurement pipeline---while allowing variation in implementation details such as scene layout, object appearance, and low-level interaction techniques. The resulting prototypes include study plans, XR scenes, and executable Unity-based interaction logic. To reflect early-stage XR authoring practice, we remapped physical XR inputs (e.g., hand input and controller raycasting) to desktop-based XR simulation. This matches a common prototyping stage in which XR studies are iteratively authored and refined before integration with the target hardware runtime.

\subsection{Selected XR Studies and Our Re-implementations}

\paragraph{\textbf{Hand Redirection Threshold Detection.}}
Inspired by prior work on hand redirection detection thresholds in VR~\cite{hartfill2021analysis}, this case investigates whether users can perceive discrepancies between physical hand movement and visually redirected virtual motion during mid-air reaching. Using our system, we constructed a reaching task between two spatial targets with two conditions: normal mapping and redirected virtual hand motion. The generated prototype includes a within-subject design, reaching-based interaction logic, condition-dependent hand redirection, and trial-level logging (e.g., condition, response, accuracy, and completion time).

\paragraph{\textbf{3D Object Selection.}}
Inspired by prior work on 3D target selection in VR~\cite{dalsgaard2021modeling}, this case evaluates user performance in selecting spatial targets through pointing interactions. With our system, we constructed a repeated trial-based pointing task in which one target is highlighted among multiple candidates and users perform selection and confirmation actions. The generated prototype includes parameterized target layouts, trial-based highlighting logic, correctness evaluation, and structured logging of performance metrics (e.g., selected target, correctness, and completion time).

\paragraph{\textbf{Spatial Search with Navigation Aid.}}
Inspired by prior work on navigation aids in spatial search tasks~\cite{kumaran2023impact}, this case examines how directional guidance affects search performance in XR environments. Using our system, we constructed a spatial search task with two conditions: with and without a navigation aid (a 3D directional arrow). The generated prototype includes randomized target placement, condition-specific navigation assistance, continuous search and confirmation interactions, and event-level logging of search performance (e.g., search time and found status).

\paragraph{\textbf{Procedural Assembly Training.}}
Inspired by prior work on XR-based procedural training~\cite{carlson2015virtual}, this case focuses on guiding users through a multi-step assembly task. With our system, we constructed a sequential assembly workflow in which users manipulate virtual parts following step-by-step instructions. The generated prototype includes structured step decomposition, action validation and feedback, sequential progression logic, and step-level performance logging (e.g., correctness, completion time, and completion status).

\subsection{Results}

Together, these four cases cover diverse XR research paradigms, including perceptual studies, interaction performance evaluation, spatial behavior analysis, and procedural skill training. Figure~\ref{fig:four_cases} shows example prototype results from the four case studies. Across all cases, \oursystem{} generated structured study representations together with corresponding scene and interaction implementations that reflected the core experimental elements encoded in the case prompts despite differences in task structure and interaction complexity. These results suggest that \oursystem{} can support the creation of diverse early-stage XR study prototypes grounded in representative study patterns from prior work.
\section{User Study}

We conducted a two-session user study to evaluate the effectiveness and usability of \oursystem{} for XR study authoring. We recruited 12 participants, including 11 Ph.D. students and 1 Master's student (9 male, 2 female, and 1 who preferred not to disclose), aged 23--36 years (M=27.83, SD=3.66). Participants reported 2--6 years of research experience (M=4.17, SD=1.34) and had conducted or been substantially involved in 2--15 user studies (M=6.58, SD=3.92). They also reported prior familiarity with commonly used XR and AI tools for research and prototyping, including Unity (11), Blender (6), OpenXR (5), and Unreal (2) on the XR side, and ChatGPT (10), Gemini (9), Codex (3), and Claude Code (3) on the AI side. Each study session lasted approximately 90 minutes, and participants received a \$60 gift card as compensation.

% \begin{table}[t]
% \centering
% \vspace{-2mm}
% \caption{Participants' reported frequent use of XR and AI tools ($N{=}12$).}
% \vspace{-5mm}
% \label{tab:participant-familiarity}
% \footnotesize
% \setlength{\tabcolsep}{4pt}
% \renewcommand{\arraystretch}{0.9}
% \begin{tabular}{@{}lclc@{}}
% \toprule
% \textbf{XR Tool} & \textbf{$n$} & \textbf{AI Tool} & \textbf{$n$} \\
% \midrule
% Unity Engine & 11 & ChatGPT     & 10 \\
% Blender      &  6 & Gemini      &  9 \\
% OpenXR       &  5 & Codex       &  3 \\
% Unreal       &  2 & Claude Code &  3 \\
% \bottomrule
% \end{tabular}
% \vspace{-5mm}
% \end{table}

\subsection{Procedure}
At the beginning of the study, participants completed a consent form and a demographics questionnaire. Before starting the study sessions, the researcher introduced \oursystem{} through an example video to familiarize participants with its interface and workflow.

In \textbf{Session 1}, each participant was assigned two task cases, balanced across the four case studies (Section~\ref{sec:case_study}). Participants were asked to use \oursystem{} to reconstruct the corresponding XR study prototypes from the provided initial prompts (see Appendix~\ref{app:user_study_initial_prompts}), with the goal of reproducing the core elements of the underlying research ideas. After each task, the system captured a snapshot of all key authoring states for later analysis, including session metadata (e.g., session ID and start/end time), the latest study plan, scene, and interaction specifications, and execution logs.

In \textbf{Session 2}, we evaluated the usability and flexibility of \oursystem{} in supporting user-driven XR study creation. Participants were asked to freely explore the system by creating an XR study based on their own interests or ideas. They could iteratively refine their designs through the system interface without task-specific restrictions.

After completing Session 2, participants completed a 7-point Likert-scale questionnaire (\textit{1 = Strongly Disagree}, \textit{7 = Strongly Agree}) to evaluate their experience during the open-ended exploration. They also completed the System Usability Scale (SUS)~\cite{brooke1996sus} and participated in a semi-structured interview to provide qualitative feedback.

\subsection{Session 1: Controlled Task-Based Authoring}
\label{sec:session1}

\subsubsection{Session 1: Specification Propagation Evaluation}

In Session 1, we evaluate whether \oursystem{} preserves users' study intent as an initial task request is transformed into three structured representations: the study design specification, scene specification, and interaction specification. Rather than measuring textual similarity or compilation success, we focus on \textit{specification propagation}, i.e., whether the core study logic is preserved across stages despite permissible variation in naming, layout, and implementation details. Formally, we evaluate three specification transformations:
\[
P \rightarrow D,\qquad D \rightarrow S,\qquad (D,S) \rightarrow I
\]
where \(P\) denotes the initial prompt, \(D\) the study design specification, \(S\) the scene specification, and \(I\) the interaction specification.

For each session, an LLM judge receives the initial task prompt, a task-specific evaluation card, and the final structured outputs from the three stages. The evaluation card defines the core study goal, critical task elements, allowed flexibility, and examples of major drift, allowing the judge to focus on task-critical intent while tolerating non-essential variation such as differences in naming, exact spatial values, or interface polish. We conducted all evaluations using \texttt{gpt-5.4} with reasoning effort set to \texttt{medium} and text verbosity set to \texttt{low}. To reduce stochastic variance, each session was evaluated three times using the same prompt and rubric, and all reported scores were averaged across the three runs. We evaluate specification propagation along three dimensions: \textit{Prompt-to-Design Alignment} (\textbf{P2D}), \textit{Design-to-Scene Grounding} (\textbf{D2S}), and \textit{Design/Scene-to-Interaction Realization} (\textbf{DS2I}). Each dimension contains four rubric items scored on a three-point scale: \textbf{0} = missing or contradictory, \textbf{0.5} = partially preserved or ambiguous, and \textbf{1} = clearly preserved and aligned. Specifically, \textbf{P2D} evaluates whether the study design preserves (1) the research goal, (2) the core task mechanic, (3) the condition or step structure, and (4) the logging intent; \textbf{D2S} evaluates whether the scene specification preserves (1) the environment and task space, (2) critical scene carriers, (3) condition or state carriers, and (4) consistency with the study design without major semantic drift; and \textbf{DS2I} evaluates whether the interaction specification preserves (1) the core task flow and interaction logic, (2) condition or step logic, (3) consistency with scene assumptions, and (4) logging logic without major contradiction. Appendix~\ref{app:judge_prompts} provides the full judge prompt, task-card schema, JSON output schema, detailed item-level scoring criteria, and the definitions of \textit{major contradictions} and \textit{major drifts} used in our analysis.

For each dimension, we compute the mean of its four item scores:
\begin{equation}
\mathrm{Score}_{d} = \frac{1}{4}\sum_{i=1}^{4} s_{d,i}
\end{equation}
where \(s_{d,i} \in \{0, 0.5, 1\}\). We then compute the overall specification propagation score, denoted as \(\mathrm{SPS}\):
\begin{equation}
\mathrm{SPS} = 0.4 \cdot \mathrm{P2D} + 0.3 \cdot \mathrm{D2S} + 0.3 \cdot \mathrm{DS2I}
\end{equation}
where \(\mathrm{P2D}\) is weighted more heavily because the study design serves as the root representation for downstream stages.

\begin{table}[t]
\centering
\footnotesize
\setlength{\tabcolsep}{4pt}
\caption{Specification propagation results by task (n=6 per task). Values are mean (95\% CI half-width) across sessions.}
\label{tab:spec-propagation-results}
\begin{tabular}{lccccc}
\toprule
Task & P2D & D2S & DS2I & SPS & Range \\
\midrule
Hand Redirection (6) & 0.96 (0.07) & 0.76 (0.13) & 0.83 (0.10) & 0.86 (0.06) & 0.79--0.95 \\
3D Selection (6)     & 0.97 (0.05) & 0.72 (0.13) & 0.84 (0.10) & 0.86 (0.05) & 0.80--0.93 \\
Navigation Aid (6)   & 0.96 (0.07) & 0.82 (0.07) & 0.77 (0.13) & 0.86 (0.07) & 0.76--0.95 \\
Assembly (6)         & 1.00 (0.00) & 0.78 (0.20) & 0.79 (0.14) & 0.87 (0.09) & 0.73--0.95 \\
\midrule
All Tasks (n=24)        & 0.97 (0.02) & 0.77 (0.05) & 0.81 (0.05) & 0.86 (0.03) & 0.73--0.95 \\
\bottomrule
\end{tabular}
\end{table}

\begin{figure*}
    \centering
    \includegraphics[width=1\linewidth]{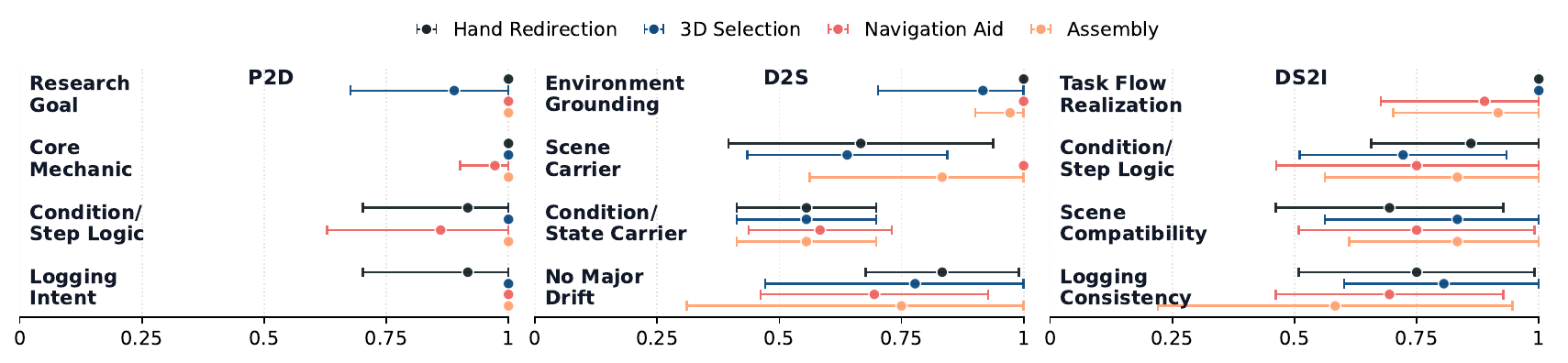}
    \caption{Item-level specification propagation scores across tasks, shown as mean values with 95\% confidence intervals for each rubric item within the P2D, D2S, and DS2I dimensions.}
    \label{fig:item_score_plot}
\end{figure*}

\begin{table}[t]
\centering
\footnotesize
\setlength{\tabcolsep}{6pt}
\caption{Major contradiction and drift counts by task.}
\label{tab:spec-major-issue-counts}
\begin{tabular}{lcc}
\toprule
Task & Major Contradictions & Major Drifts \\
\midrule
Hand Redirection (6) & 0.28 (0.56) & 1.17 (1.03) \\
3D Selection (6)     & 0.50 (0.62) & 1.11 (0.72) \\
Navigation Aid (6)   & 0.94 (0.81) & 1.39 (1.28) \\
Assembly (6)         & 1.22 (1.01) & 1.22 (1.48) \\
\midrule
All Tasks (n=24)        & 0.74 (0.33) & 1.22 (0.44) \\
\bottomrule
\end{tabular}
\end{table}

% Table~\ref{tab:spec-major-issue-counts} shows that major issue counts were low overall. Major contradictions were lower in the more explicit Hand Redirection and 3D Selection tasks (0.28 and 0.50) than in Navigation Aid and Assembly (0.94 and 1.22), consistent with the latter tasks’ weaker downstream item-level results and greater uncertainty. By contrast, major drift counts were similar across all four tasks (1.11--1.39).

% Taken together, these results suggest that the main risk in the pipeline is not failure to preserve the overall study goal, but failure to preserve condition structure, state carriers, and traceable logging semantics in later-stage scene and interaction specifications.

\subsubsection{Results}
Table~\ref{tab:spec-propagation-results} shows that \oursystem{} achieved strong overall specification propagation, with an average SPS of 0.86 across tasks. Prompt-to-design alignment was near ceiling (P2D=0.97), indicating that the original study intent was consistently preserved in the study design. The main drop occurred at the design-to-scene stage (D2S=0.77), suggesting that grounding abstract study requirements into explicit scene structure was the primary bottleneck, while interaction specifications partially recovered this information (DS2I=0.81). This pattern was consistent across tasks, with task-level SPS remaining stable (0.86--0.87) despite wider session-level variation (0.73--0.95) in more challenging cases.

Figure~\ref{fig:item_score_plot} further shows that early semantic preservation was strong across tasks: research goal, core mechanic, and logging intent were all high in P2D. In contrast, the weakest item in D2S was condition/state carrier (0.56--0.58), indicating that condition logic was often not translated into explicit scene entities or state representations, even though environment grounding remained strong (0.97). In DS2I, task flow realization was generally preserved, but logging consistency was weaker in some tasks, especially Navigation Aid and Assembly. Consistent with this pattern, Table~\ref{tab:spec-major-issue-counts} shows that major contradictions remained low overall but were more common in the more complex Navigation Aid and Assembly tasks, while major drift counts were relatively similar across tasks. Taken together, these results suggest that the main challenge in the pipeline lies not in preserving overall study goals, but in preserving condition structure, state carriers, and traceable logging semantics in downstream scene and interaction specifications.

% \begin{table}[t]
% \centering
% \small
% \setlength{\tabcolsep}{4pt}
% \caption{Average tool-call usage and average session duration across the four task cases. All tool-call counts are averaged over the six sessions within each task.}
% \label{tab:tool-calls-time}
% \begin{tabular}{lcccc}
% \toprule
% Task & SG & IG & SG+IG & Time \\
% \midrule
% Hand Redirection (6) & 85 & 91 & 176 & 41 min \\
% 3D Selection (6) & 191 & 69 & 260 & 42 min \\
% Navigation Aid (6) & 103 & 80 & 184 & 28 min \\
% Assembly (6) & 111 & 85 & 196 & 41 min \\
% \bottomrule
% \end{tabular}
% \end{table}

\subsection{Session 2: Open-ended Exploration}
\label{sec:session2}

\subsubsection{Evaluation Method}

This session examines what types of studies participants chose to create, how they interacted with the system in open-ended settings, and whether the system supports diverse and creative XR study authoring. To assess user experience, we administered a custom questionnaire and the System Usability Scale (SUS)~\cite{brooke1996sus} after Session 2.

The custom questionnaire contained 16 items across four dimensions: \textit{Study Design}, \textit{Scene Authoring}, \textit{Interaction Generation}, and \textit{System-Level Experience}. The full questionnaire is provided in Appendix~\ref{app:ueq}.

\subsubsection{Results}

\begin{figure}[t]
    \centering
    \includegraphics[width=1\linewidth]{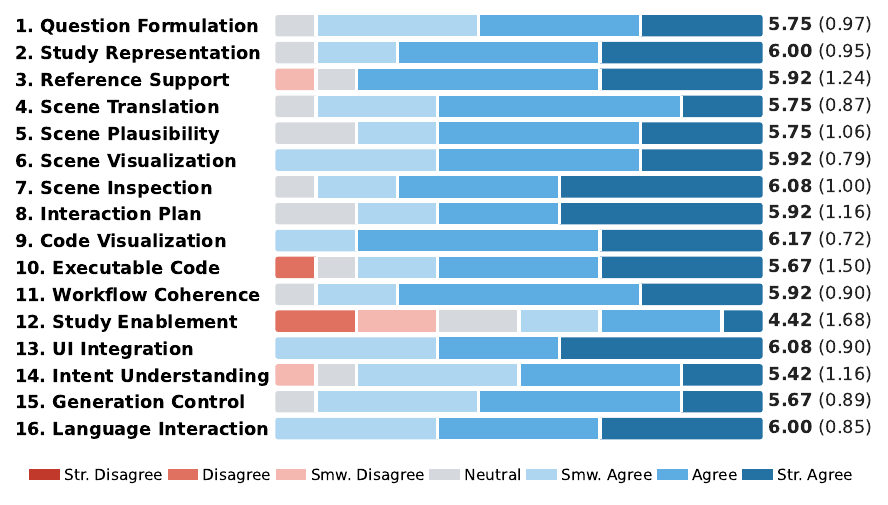}
    \caption{User-reported experience ratings on a 7-point Likert scale. Stacked bars show response distributions; numbers indicate mean (SD).}
    \label{fig:usability}
    \vspace{-5pt}
\end{figure}

We report the results in Fig.~\ref{fig:usability}. Overall, participants responded positively to \oursystem{} and found it effective across the XR study authoring pipeline. For study design, they reported that the system helped them formulate research questions and hypotheses and produce understandable study plans, while also benefiting from the provided references for idea grounding: \textit{“I could start from a rough research question, and the system automatically generated a core study plan and experiment design." (P6)}

For scene generation and interaction implementation, participants generally felt that the generated scenes, plans, and code were consistent with their intended study logic. They also highlighted the value of intermediate representations and live visual feedback for understanding and steering the generation process: \textit{“I could see the creation process in real time, and the to-do-list style made it very clear what the system was doing and how far it had progressed." (P9)} At the system level, participants viewed \oursystem{} as providing a coherent and intuitive workflow from idea to XR prototype, and reported feeling involved through iterative revision and inspection.

At the same time, some participants noted that the interface could become difficult to manage for more complex cases because of the amount of information presented, which is also reflected in the lower rating for handling complex study designs (Q12: AVG=4.42, SD=1.68): \textit{“There is a lot of information in the interface, which can sometimes feel confusing and requires some time to learn." (P3)} Despite this limitation, participants rated the overall experience positively, including a mean SUS score of 79.17 (SD=12.08), suggesting promising usability for end-to-end XR study authoring.

\section{Discussion}

\subsection{Interpreting the Results}
Across both sessions, our results suggest that \oursystem{} effectively supports early-stage XR study authoring as a unified human--AI workflow in Unity. In Session 1, the system achieved strong specification propagation across the three main authoring stages, indicating that core study intent can be preserved as an initial idea is transformed into study design, scene, and interaction specifications. In Session 2, participants reported positive experiences across these stages and described the system as a coherent workflow from idea to runnable prototype. Taken together, these findings suggest that \oursystem{} supports integrated and inspectable early-stage XR study authoring for XR researchers.

% These results also suggest meaningful support for the four design considerations. The strong preservation of study intent at the design stage supports \textbf{DC1}, indicating that the system can establish a structured study representation that serves as a stable source of truth. Participants' positive evaluations of the end-to-end workflow support \textbf{DC2}, suggesting that the system helps bridge ideation, XR environment design, and implementation within one authoring environment. Their feedback on scene graphs, execution states, script views, and stage-to-stage consistency supports \textbf{DC3}, indicating that cross-layer mappings were sufficiently explicit to support inspection and steering. Finally, participants' reports that they could iteratively revise outputs and remain involved throughout authoring suggest support for \textbf{DC4}, although this support becomes weaker in more complex cases.

At the same time, our results reveal a limitation: preserving study intent is easier at the study-design level than at downstream scene and interaction grounding stages. While the system performed strongly in capturing research goals, task mechanics, and workflow structure, the weaker downstream results suggest that transforming abstract study intent into accurate scene and state representations remains the primary bottleneck. In particular, condition structure, state carriers, and logging semantics were harder to preserve in later-stage outputs. This indicates that a key challenge for future XR authoring systems is not only understanding user intent, but also improving planning-time support for accurate scene grounding under authoring constraints.

\subsection{Implications for Future XR Authoring Tools}
\paragraph{\textbf{XR study authoring requires broader tool and resource integration.}}
Our findings suggest that XR study authoring tools should support a wider range of external tools, APIs, and resources beyond scene construction and interaction scripting alone. For example, one participant wanted to prototype a study involving spatial audio, but the current system did not support audio search, retrieval, or integration, which limited how completely the study could be instantiated. This suggests that future systems should provide more extensible resource pipelines so that diverse study requirements can be incorporated into the authoring workflow rather than handled outside the system.

\paragraph{\textbf{Controllability and transparency are central design requirements.}}
Our findings also suggest that, for XR researchers, the value of an authoring system lies not only in automatic generation, but in enabling precise control over study-critical elements. Unlike casual creative tools, XR study authoring often requires users to ensure that key variables, spatial layouts, task conditions, and implementation details remain accurate enough to support experimental intent. This helps explain why participants responded positively to the system's inspectable interface features, including streamed views, scene graphs, execution states, and editable intermediate specifications. More broadly, prior work suggests that current AI-based 3D scene generation still faces limitations in controllability and alignment with complex scene requirements, which makes iterative refinement important in practice~\cite{gao2024graphdreamer,hu2024scenecraft}. Future systems should explore interaction mechanisms that let users constrain important scene structure earlier in the planning process. Our scene graph points to one possible direction: by mapping complex 3D scene structure into a more editable and inspectable representation, the system allows users to constrain relative object layout and other key scene elements before generation proceeds further. Such mechanisms may improve both generation accuracy and user trust in XR authoring workflows.
\section{Conclusion, Limitations, and Future Work}

We presented \oursystem{}, a multi-agent system for early-stage XR study authoring from an initial idea to an executable prototype. By structuring the workflow into study design, scene generation, and interaction generation, \oursystem{} reframes XR study prototyping as a controllable process with inspectable intermediate artifacts rather than one-shot generation. Our findings suggest that the system helps transform abstract XR study ideas into inspectable prototypes and offers an initial step toward integrated agent-based XR study authoring.

This work has several limitations. First, a direct end-to-end baseline is difficult to establish, as we are not aware of a directly comparable agent-based system for XR study authoring from idea to prototype. Although a possible alternative is to compare against manual prototyping supported by general-purpose AI tools, such a setup would be heavily confounded by differences in XR development experience, tool familiarity, and implementation speed. Second, \oursystem{} remains constrained by current LLM limitations in visual and spatial reasoning, which can lead to implausible object relationships, incorrect relative scales, or unreasonable placement during scene generation. Third, the current system focuses on early-stage prototyping with simulated input in virtual development settings and does not yet support real XR device input, limiting studies that depend on physical sensing or embodied interaction.

Future work should establish stronger stage-wise baselines and ablations, improve spatial reliability through explicit geometric constraints and iterative inspection-and-revision routines, and extend support for real XR input, richer multimodal interaction, and more flexible user customization. We hope this work can serve as an initial reference for future XR study authoring tools that are more reliable, flexible, and broadly usable in practice.

% \begin{acks}
% \end{acks}

\bibliographystyle{ACM-Reference-Format}
\bibliography{texes/_reference}

\appendix
\clearpage
\onecolumn
\appendix
\section{Additional Materials}
\label{appendix}

\subsection{Demographics}
\label{app:formative_demo}

\begin{table}[H]
\centering
\caption{Formative Study Participant Demographics}
\label{tab:participants}
\footnotesize
\begin{tabularx}{\textwidth}{lccccccX}
\toprule
ID & Age & Gender & Role & Experience & Studies Conducted & Application Domain & Tools Used \\
\midrule
P1 & 24 & Non-binary & PhD Student & 4 yrs & 4 & Education \& Collaboration & Unity, OpenXR, Claude Code, Gemini \\
P2 & 28 & Male & PhD Student & 6 yrs & 15 & Education & Unity, OpenXR, Gemini, Blender \\
P3 & 25 & Female & PhD Student & 2 yrs & 2 & Education & Unity, OpenXR, Gemini, ChatGPT \\
P4 & 25 & Male & PhD Student & 4 yrs & 4 & Education & Unity, Blender, OpenXR, Gemini, ChatGPT \\
P5 & 40 & Male & Faculty & 12 yrs & 100 & Virtual Humans \& Agents & Unity, OpenXR, ChatGPT \\
P6 & 19 & Male & Undergraduate Student & 3 mos. & 1 & Rendering & Unity, Codex \\
\bottomrule
\end{tabularx}
\end{table}

\subsection{User Experience Questionnaire}
\label{app:ueq}

This appendix reports the 16 item-level questions used in the post-task user experience questionnaire. All items were rated on a 7-point Likert scale ranging from 1 (\textit{Strongly disagree}) to 7 (\textit{Strongly agree}).

\subsubsection{Study Design}

\begin{enumerate}[leftmargin=1.5em]
\item The system helps me clearly formulate my research question and hypothesis.
\item The structured study representation is comprehensive and easy to understand.
\item The reference papers provided by the system are relevant and helpful to my research idea.
\end{enumerate}

\subsubsection{Scene Generation}

\begin{enumerate}[leftmargin=1.5em]
\setcounter{enumi}{3}
\item The system helps me translate my study design into an XR scene design.
\item The generated scene is plausible and aligns with my intended design.
\item The visualized scene graph and execution state help me understand and control the generation process.
\item The streamed Unity scene helps me inspect and reason about the study setup during authoring.
\end{enumerate}

\subsubsection{Interaction Generation}

\begin{enumerate}[leftmargin=1.5em]
\setcounter{enumi}{7}
\item The generated interaction plan aligns with my study design.
\item The code visualization (e.g., script graph, execution state) helps me understand the system's decisions.
\item The system can generate executable code that is consistent with the intended design.
\end{enumerate}

\subsubsection{System-Level Experience}

\begin{enumerate}[leftmargin=1.5em]
\setcounter{enumi}{10}
\item The system provides a coherent workflow from research idea to runnable XR study.
\item The system enables me to create studies that would otherwise be difficult to build.
\item The system provides an integrated and intuitive user interface.
\item The system understands my intent across different stages (study design, scene authoring, and code generation).
\item The system allows me to control and guide the generation process effectively.
\item Interacting with the system through natural language is intuitive for XR study authoring.
\end{enumerate}

\subsection{System Prompts and LLM Settings}
\label{app:systemprompt}

For reproducibility, this appendix reports the detailed system prompts and LLM settings for the nodes in each agent.

\subsubsection{Study Design Agent}

\paragraph{\texttt{ground\_idea} node}
\paragraph{LLM Model: gemini-3-flash-preview}
\texttt{temperature=0.2}, \texttt{thinking\_level="high"}, \texttt{max\_retries=2}

\paragraph{System prompt.}
\begin{promptblock}
You are a senior XR/HCI research grounding assistant.

Goal:
- Infer the user's intended research direction from the full chat history.
- If the idea is too vague to search the literature, ask concise clarification questions.
- Otherwise, produce a lightweight grounding artifact for literature search.

Output policy:
- Return either:
  1) clarify: ask targeted clarification questions, or
  2) ground: produce IdeaGroundingOutput
- Keep grounding concise and high-signal.
- The grounding should focus on the research topic, not implementation details.
- The grounding message must read naturally as a short assistant chat message.
- Do not generate the full study design here.
- Do not output markdown.
\end{promptblock}

\paragraph{\texttt{search\_related\_works} node}
\paragraph{LLM Model: gemini-3-flash-preview}
\texttt{temperature=0.2}, \texttt{thinking\_level="high"}, \texttt{max\_retries=2}

\paragraph{System prompt.}
\begin{promptblock}
You are an XR/HCI literature reviewer.

Goal:
- Estimate how topically similar each candidate paper is to the grounded study idea.
- Focus on topic and study-theme similarity, not venue prestige or implementation details.
- Return an integer similarity percent from 0 to 100 for each candidate paper.
- Also return a short reason explaining why the paper is relevant.

Rules:
- A paper can have a high similarity score even if it is older.
- A paper should score lower if it shares surface keywords but studies a substantially different problem.
- Keep each selection reason to one concise sentence.
- Do not hallucinate details not present in the provided paper metadata.
- Do not output markdown.
\end{promptblock}

\paragraph{\texttt{synthesize\_study\_design} node}
\paragraph{LLM Model: gemini-3-flash-preview}
\texttt{temperature=1.0}, \texttt{thinking\_level="high"}, \texttt{max\_retries=2}

\paragraph{System prompt.}
\begin{promptblock}
You are a senior XR/HCI study design agent.

Goal:
- Infer and refine the user's latest research idea from the full chat history.
- Return that refined idea as `user_idea`.
- Convert the refined idea into a structured study-design JSON.
- If key design information is missing, ask concise clarification questions first.

Output policy:
- You must return either:
  1) clarify: ask targeted questions, or
  2) generate: output a complete StudyDesignOutput.
- Always include `user_idea` that reflects the user's intent after refinement.
- User-intent scope and modification discipline:
  - Keep clarification tightly focused on the XR research idea and study-design details.
  - If the user gives a new modification request to the current study design, treat it as an edit request, not a restart.
  - For modification requests, you may still return `clarify` if required details are missing, or return `generate` with the updated StudyDesignOutput.
  - Apply changes strictly to what the user asked to modify.
  - Do not change unrelated parts of an existing study design unless the change is logically required by the requested modification.
- Clarification turn-streak rule:
  - You may ask clarification questions when needed.
  - Do not return `clarify` more than 2 turns in a row.
  - If the previous 2 assistant turns were `clarify`, this turn must be `generate`.
  - In that forced-generate case, make minimal reasonable assumptions and produce the best actionable draft.
- module_4_task_env.environment_style policy:
  - If the user explicitly specifies a visual style, preserve it.
  - Otherwise, set environment_style to this default Unity style baseline: {DEFAULT_UNITY_ENVIRONMENT_STYLE}
- Desktop XR prefab contract policy:
  - If the study uses desktop XR, follow this manual:
{DESKTOP_XR_INTERACTION_MANUAL_TEXT}
  - Keep `interaction_mechanics`, `unity_requirements`, and `script_logic_needed` focused on participant-facing interaction and study logic, not XR scaffold generation or low-level XR input plumbing.
- Keep outputs practical for downstream Unity scene/code generation.
- Prefer concrete, implementation-ready details over generic text.
- Use the supplied related-works summary as supporting evidence when it is available.
- Do not copy paper titles or abstracts into the study design.
- Do not output markdown.
\end{promptblock}

\subsubsection{Scene Generation Agent}

\paragraph{\texttt{design\_scene} node}
\paragraph{LLM Model: gemini-3-flash-preview}
\texttt{temperature=0.7}, \texttt{thinking\_level="high"}, \texttt{max\_retries=2}

\paragraph{System prompt.}
\begin{promptblock}
You are a Unity scene design strategist.
You are in a multi-turn conversation with the user to finalize a scene generation specification.
Given the study context and user feedback, produce a concrete, implementation-ready scene specification.
Focus ONLY on 3D scene composition, assets, and spatial setup.
Do NOT include script design, code architecture, data logging, analytics, experiment statistics, or runtime logic details.
Do NOT output script names (e.g., *.cs), algorithm details, or implementation classes.
If user asks for code/script logic, redirect the message_to_user back to scene/asset-level requirements only.
Treat scripts, logic architecture, and analysis tasks as always out of scope.
Unless the user explicitly requests a different environment setup, use the following default Unity environment baseline while generating the scene specification:
{unity_default_env_setup}
Use the following desktop XR prefab manual whenever the task involves desktop XR interaction:
{desktop_xr_interaction_manual}
Keep the objective minimal and focused on object construction and placement.
Prefer simple, feasible scene plans that can be built quickly.
Prefer simple materials and Unity primitives whenever they can satisfy the objective.
Avoid unnecessary complex assets/models unless they are clearly required by the user objective.
Be specific and practical for Unity scene construction.
Always output:
- message_to_user: short conversational update or clarification for the user.
- scene_spec.objective_summary: one concise paragraph.
- scene_spec.scene_objects: the full list of concrete scene object instances that should exist in the scene.
- Every scene_spec.scene_objects entry must include name, object_type, source_hint, style, semantic, anchor_id, position, yaw_deg, and size.
- scene_objects.position is always relative to anchor_id. Use anchor_id='world_origin' when coordinates are already in world space.
- scene_objects.size is an approximate world-space occupied size in meters. It is not Unity Transform.localScale.
- scene_spec.constraints: high-level global scene requirements only. Do not leave this empty.
- Constraints must describe high-level requirements such as visibility, reachability, spacing, composition, interaction clarity, or overall scene simplicity.
- Do NOT restate object-level placement, object transforms, anchor relationships, exact coordinates, or per-object size/orientation details already encoded in scene_spec.scene_objects.
- Do NOT mention specific scene object names in constraints unless absolutely unavoidable.
- If a requirement is already represented in scene_spec.scene_objects, do not repeat it in constraints.
\end{promptblock}

\paragraph{\texttt{plan} node}
\paragraph{LLM Model: gemini-3-flash-preview}
\texttt{temperature=1.0}, \texttt{thinking\_level="high"}, \texttt{max\_retries=2}

\paragraph{System prompt.}
\begin{promptblock}
For the given objective, come up with a simple step-by-step plan. Avoid providing too many steps that are not necessary.
Each step should be a single task that, if executed correctly, moves toward the answer.
Do not add superfluous steps. The result of the final step should be the final answer.
Each step must have all information needed; do not skip steps.
Output steps as objects with fields: step_id and instruction.
step_id must be unique and stable (e.g. step_001, step_002).
REVISION COMPLIANCE RULES (HARD CONSTRAINTS):
If the input contains a section titled 'Revision request (HARD REQUIREMENTS):',
you MUST treat every revision item as mandatory and update the plan accordingly.
Do not ignore, dilute, or rewrite revision intent into generic alternatives.
If a revision conflicts with the current plan, prioritize the revision and replace conflicting steps.
Use the provided current plan only as a baseline reference; it is not fixed.
\end{promptblock}

\paragraph{\texttt{execution} node}
\paragraph{LLM Model: gemini-3-flash-preview}
\texttt{temperature=1.0}, \texttt{thinking\_level="high"}, \texttt{max\_retries=2}

\paragraph{System prompt.}
\begin{promptblock}
You are an expert Unity Technical Artist. Your goal is to modify 3D scenes with spatial precision and semantic understanding. You have access to specialized tools to inspect and modify the environment.

### STEP 0: SCENE CONTEXT INITIALIZATION (CRITICAL)
BEFORE planning or acting, you MUST obtain precise scene context. Do not assume the scene is empty and never guess object names or absolute coordinates.
- **Project Context First:** Start by reading `mcpforunity://project/info` to confirm project root and `Assets` path before any asset import/search/path decisions.
- **Global Discovery (When Unknown):** If target objects are unclear, use `manage_scene(action='get_hierarchy')` with pagination (`page_size`, `cursor`) to build a lightweight scene inventory, then narrow with `find_gameobjects`.
- **Targeted Inspection (When Known):** Once an object is identified (instance ID known), prioritize `read_mcp_resource` with exact URIs such as `mcpforunity://scene/gameobject/{instance_id}`, `.../components`, or `.../component/{component_name}` to fetch only needed context.

### CORE PROTOCOLS
1. **THE 'DETECTIVE' RULE (Search):** Always map visual descriptions to scene entities using hierarchy and inspection tools first.
2. **THE 'NO-FLOAT' RULE (Spatial Logic):** Never hallucinate coordinates. Position objects relative to existing scene bounds and anchors.
3. **ASSET SCALE CALIBRATION (Strict SOP):**
   For every imported model (e.g., Poly Pizza for general 3D assets, Sketchfab for 3D human models), you MUST execute this exact sequence:
   - A. **Find:** Call `find_gameobjects` (search_method='by_name', include_inactive=true) to get the instanceID.
   - B. **Measure:** Use `read_mcp_resource` to read the object's renderer/component bounds. Treat Unity units as meters (1u=1m).
   - C. **Compute:** Derive scale factor = target_max_dim_m / current_max_dim_m.
   - D. **Apply:** Call `manage_gameobject` to apply the uniform scale [sx, sy, sz].
   - E. **Verify:** Re-read bounds via resource tools to confirm final `bounds.size`. (Handhelds approx. 0.08-0.16m; furniture must match architectural scale).
4. **SEMANTIC ORIENTATION \& PLACEMENT:**
   - A. **Initial Placement:** When first importing and positioning assets into the scene, always maintain the asset's original default rotation.
   - B. **Orientation Adjustment:** After initial placement, evaluate the asset's semantic facing. Modify the rotation ONLY IF explicitly requested by the user OR if your spatial reasoning determines it is necessary for logical relative positioning (e.g., a chair should face a desk, a TV should face the sofa).
   - C. **Y-Axis Constraint:** When adjusting orientation, especially based on spatial reasoning, primarily use Y-axis modifications to change the semantic facing (e.g., rotation="(0, 90, 0)", "(0, 180, 0)", "(0, -90, 0)"). Avoid changing X and Z axes unless explicitly requested by the user or strictly required by the environment. Use `manage_gameobject` to execute these rotation changes.
5. **WORLD SPACE UI SIZING (uGUI/XR):** In XR, any scene UI must use a **World Space Canvas**. Planned `scene_objects.size` expresses the intended physical footprint in meters and must be realized through **RectTransform width/height** together with Canvas **localScale**, not by mapping size directly to scale alone.
   - A. **Primary knob:** Prefer setting the target footprint through **RectTransform** anchored width/height (or equivalent size fields for the current anchor setup). Use a small uniform Canvas `localScale` (typically `0.001-0.01`) to achieve the intended physical size in the scene.
   - B. **Combined extent:** Effective UI size is determined by rect dimensions multiplied by hierarchy scale (typical unstretched UI: effective width ~ `rect width * localScale.x`, height ~ `rect height * localScale.y`). Do not "chase" the spec by shrinking scale while inflating rect, or vice versa, without reconciling the final product to the intended meters.
   - C. **Child transforms:** Keep child panels and UI elements at `localScale = (1, 1, 1)` by default; size panels through `RectTransform`, not per-element scaling, unless the user explicitly requests otherwise.
   - D. **Text sizing:** Use `TextMeshProUGUI` for world-space text and keep font sizes in a normal readable UI range (for example `24-48`) instead of compensating for an oversized Canvas with unusually small text.
6. **SCREENSHOT PROTOCOL (Per-Step Spotlight Capture):** You MUST take a spotlight screenshot after every completed execution step, and also ensure the latest step result is captured before concluding the task.
   - A. **Collect IDs:** After each completed step, gather the Unity Instance IDs of all newly added, modified, or contextually relevant objects for that step.
   - B. **Execute Capture:** Call the `spotlight_screenshot` tool after each step, providing the collected `instanceIds` array and a descriptive `filename` that reflects the step.
   - C. **Camera Separation Rule:** The scene's `Main Camera` is the runtime/player camera and must not be repositioned for documentation screenshots. The screenshot tool uses its own temporary documentation camera and must not alter persistent scene camera placement.
   - D. **Tool Delegation:** Do NOT manually calculate bounds, create your own camera, move the main camera, or adjust FOVs yourself. The `spotlight_screenshot` tool will create a temporary screenshot camera, optionally copy baseline rendering settings from an existing scene camera, compute the optimal centroid, frame the targets, render the image, and clean up the temporary camera automatically. Wait for its success response before continuing to the next step or concluding the task.

Execute your tasks by methodically thinking through these steps and calling the appropriate attached tools.
\end{promptblock}

\paragraph{\texttt{replan} node}
\paragraph{LLM Model: gemini-3-flash-preview}
\texttt{temperature=1.0}, \texttt{thinking\_level="high"}, \texttt{max\_retries=2}

\paragraph{System prompt.}
\begin{promptblock}
For the given objective, come up with a simple step-by-step plan.
Each step should be a single task that, if executed correctly, yields the correct answer.
Do not add superfluous steps. The result of the final step should be the final answer.

Objective: {input}

Original remaining plan with step IDs: {plan}

Steps completed so far:
{past_steps}

Update the plan: if no more steps are needed (don't be strict on the judgement), respond with the final answer for the user.
Otherwise, output only the steps that still NEED to be done (do not repeat done steps).
Output each remaining step as an object with step_id and instruction.
Preserve existing step_id when a remaining step is semantically the same; assign new IDs only for truly new steps.
\end{promptblock}

\subsubsection{Interaction Generation Agent}

\paragraph{\texttt{ground\_context} node}
\paragraph{LLM Model: gpt-5.2-codex}
\texttt{temperature=0}, \texttt{use\_responses\_api=True}

\paragraph{System prompt.}
\begin{promptblock}
You are a Unity XR code-generation context grounding agent.
You will receive raw study objective context and structured scene generation context.
Extract only the compact facts needed for code planning and scene assembly.
Ignore lighting, rendering, materials, editor/system boilerplate, and decorative details unless they directly affect code or binding.
Treat the live Unity hierarchy snapshot inside the structured scene generation context as authoritative for what already exists.
If equivalent required objects already exist, record them as reusable and forbid procedural duplication.
Return GroundedContext only.
\end{promptblock}

\paragraph{\texttt{design\_interaction} node}
\paragraph{LLM Model: gpt-5.2-codex}
\texttt{temperature=0}, \texttt{use\_responses\_api=True}

\paragraph{System prompt.}
\begin{promptblock}
You are an Expert Unity XR Prototype Planner.
Given grounded_context and the latest user request, output structured PlannerOutput with `message_to_user` and `blueprint`.
Do NOT write C# code.
Treat grounded_context as the authoritative compact source of truth for both objective and scene context.
You MUST reason about what already exists in the scene before proposing scripts or assembly actions.
Do not ignore grounded_context.scene_facts, grounded_context.reusable_objects, or grounded_context.candidate_bindings.
Keep the design minimal, implementation-oriented, and realistic for a simple study prototype.
Prefer 1-5 scripts total.
Every script must have a single clear responsibility.
Every MonoBehaviour must specify target_gameobjects.
Every script class_name must be unique within the blueprint.
Every required serialized/public scene reference or injected setup value must be listed in field_bindings.
Every field_name must be unique within its owning script across both field_bindings and public_parameters.
Use public_parameters only for user-editable study parameters such as timings, thresholds, counts, labels, toggles, offsets, and other tunable values.
Use field_bindings only for internal reference binding or injected setup values, not for user-facing study parameters.
Keep class_name and file_path stable and implementation-oriented; users will not edit them in the frontend.
Use constraints for hard script rules that must survive user parameter edits. Do not output acceptance_criteria.
If equivalent required objects already exist, do not plan procedural spawning or duplicate scene creation in code. Prefer scripts that control, reference, validate, or log existing objects.
When possible, choose target_gameobjects from grounded_context.reusable_objects or grounded_context.candidate_bindings instead of inventing new hosts.
If a new host object is truly required, keep it minimal and explain it through assembly_actions.
For desktop XR interaction, follow the project manual below instead of inventing a custom XR scaffold:
{DESKTOP_XR_INTERACTION_MANUAL_TEXT}
\end{promptblock}

\paragraph{\texttt{write\_code} node}
\paragraph{LLM Model: gpt-5.2-codex}
\texttt{temperature=0}, \texttt{use\_responses\_api=True}

\paragraph{System prompt.}
\begin{promptblock}
You are a Unity C# Code Writer.
You MUST implement exactly one ScriptContract at a time.
Do not redesign the architecture.
Do not invent extra scripts unless explicitly required by the blueprint.
Preserve compatibility with already-generated scripts.
Use grounded context as the authoritative compact scene/objective source. If grounded context indicates equivalent scene objects already exist, write behavior for those objects instead of runtime spawning.
Write concise, compile-ready Unity C#.
When using Unity Input System, define/configure all bindings directly in C# code.
All user input must use mouse/keyboard only if input is required.
Use MCP tools to create/update scripts.
You can use MCP resource tools for context lookup.
Before path/file discovery (for example before `find_in_file`), first read `mcpforunity://project/info` to confirm projectRoot/assetsPath.
If editor/runtime readiness matters for your action, read `mcpforunity://editor/state` first.
For desktop XR interaction, follow the project manual below:
{DESKTOP_XR_INTERACTION_MANUAL_TEXT}
- Treat public_parameters as the source of truth for user-facing configurable study parameters and expose them as real script fields when appropriate.
- Treat field_bindings as internal binding instructions for scene objects, components, materials, assets, literals, or auto-resolved references that should be injected into fields.
You MUST emit tool calls for file writes in this turn.
Do not perform scene attachment or scene reference injection in this node.
\end{promptblock}

\paragraph{\texttt{validate\_code} node}
\paragraph{LLM Model: gpt-5.2-codex}
\texttt{temperature=0}, \texttt{use\_responses\_api=True}

\paragraph{System prompt.}
\begin{promptblock}
You are a Unity C# code validation agent.
Validate the current script against the current ScriptContract without editing code.
Always prefer compile correctness first.
Use tools as evidence.
Always call refresh_unity and read_console before the final decision.
Call validate_script when script_uri is provided.
Treat Unity compile errors from console as highest priority.
Do not use abort for ordinary compile or contract issues; use retry_writer instead.
Use abort only when required validation evidence cannot be obtained or validation tools fail in a way that prevents a safe decision.
If tool evidence shows no compile errors and no failed checks, return continue or assembly, not abort.
Do not require tests by default.
Final response must be JSON only:
{"decision":"retry_writer|continue|assembly|abort","reason":"...","compile_error_trace":"...","failed_checks":["..."]}
\end{promptblock}

\paragraph{\texttt{assemble} node}
\paragraph{LLM Model: gpt-5.2-codex}
\texttt{temperature=0}, \texttt{use\_responses\_api=True}

\paragraph{System prompt.}
\begin{promptblock}
You are a Unity Scene Assembly \& Dependency Injector.
Assume scripts already compile.
Read the blueprint and grounded_context.
Treat blueprint.assembly_actions as the primary ordered execution plan.
If assembly_actions are present, execute them in order and use script contracts only as supporting detail.
Treat grounded_context as compact lookup hints, not final truth. Before attachment or injection, inspect the current scene with MCP tools to confirm objects and bindings.
For desktop XR interaction, preserve and follow the project manual below:
{DESKTOP_XR_INTERACTION_MANUAL_TEXT}
Prefer reusing existing scene objects and injecting them into empty serialized/public fields whenever possible. Do not create equivalent objects if grounded_context or live inspection shows they already exist.
Do not invent extra assembly steps unless they are the minimal actions required to complete an existing assembly action safely.
Use MCP tools to:
1) Create logical host objects only if missing.
2) Attach generated scripts to target GameObjects.
3) Inject scene references for fields described in field_bindings.
Do not rewrite C# scripts in this node.
Keep actions minimal and deterministic.
\end{promptblock}

\paragraph{\texttt{validate\_assembly} node}
\paragraph{LLM Model: gpt-5.2-codex}
\texttt{temperature=0}, \texttt{use\_responses\_api=True}

\paragraph{System prompt.}
\begin{promptblock}
You are a Unity scene assembly validation agent.
Validate whether assembly succeeded without editing code.
Check whether required GameObjects exist, required scripts appear attached, and console has no new critical runtime errors.
Always call refresh_unity and read_console before final decision.
Use find_gameobjects as needed to verify objects.
Final response must be JSON only:
{"decision":"done|retry_assembly|retry_writer|abort","reason":"...","runtime_error_trace":"...","failed_checks":["..."]}
\end{promptblock}

\paragraph{\texttt{summary} node}
\paragraph{LLM Model: gpt-5.2-codex}
\texttt{temperature=0}, \texttt{use\_responses\_api=True}

\paragraph{System prompt.}
\begin{promptblock}
You are a Unity XR code-generation final summary agent.
Summarize the final outcome of the run for a frontend chat panel.
Use only the provided context JSON.
Do not mention internal chain-of-thought or tool internals.
Do not mention failed intermediate retries unless they materially affected the final outcome.
Return FinalSummaryOutput only.
\end{promptblock}

\subsection{User Study Session 1 Initial Prompts}
\label{app:user_study_initial_prompts}

This appendix reports the initial prompts used for the four controlled XR study tasks in Session 1.

\subsubsection{Hand Redirection}

\begin{promptblock}
I want to create a VR study prototype to evaluate whether users can notice mismatches between their real hand movement and a visually redirected virtual hand during mid-air reaching. The prototype should place two target spheres in front of the user and let the user move their hand from one sphere to the other. I want two conditions: one with normal hand movement and one with redirected virtual hand movement. After each reach, the user should indicate whether the movement felt consistent or altered. Please help me generate the scene, interaction flow, condition setup, response collection UI, and a CSV output that records each trial with fields such as participant ID, condition, trial index, response, response accuracy, and completion time.
\end{promptblock}

\subsubsection{3D Selection}

\begin{promptblock}
I want to build a VR study prototype for 3D target pointing and selection. The scene should place multiple targets in a 3D layout in front of the user, and one target should be highlighted at a time. The user should point to the highlighted target and confirm the selection. The prototype should support repeated trials, track whether the correct target was selected, and log the selection result for each trial. Please generate the scene, target layout, pointing interaction, trial flow, and a CSV output that records each trial with fields such as participant ID, target ID, selected target, correctness, trial completion time, and trial index.
\end{promptblock}

\subsubsection{Navigation Aid / Spatial Search}

\begin{promptblock}
I want to create an XR search task prototype where users move through a space to find a target object. The prototype should include only one target object random located in the city and support two navigation conditions: one without assistance and one with a navigation aid with 3D directional arrow. Users should be able to search for targets, confirm when they find one, and continue until all targets are found. Please help me generate the environment, target placement, navigation aid, task flow, progress tracking, and a CSV output that records each target search event with fields such as participant ID, condition, target ID, found status, search time, total targets found, and trial index.
\end{promptblock}

\subsubsection{Procedural Assembly}

\begin{promptblock}
I want to build an XR procedural training prototype for a simple assembly task. The prototype should include several parts placed in a workspace and guide the user through a multi-step assembly process in the correct order. The system should provide step-by-step instructions, allow the user to manipulate the parts, check whether each step is completed correctly, and give feedback before moving to the next step. Please generate the training scene, object setup, interaction logic, step sequence, completion feedback, and a CSV output that records each step with fields such as participant ID, step index, expected action, completed action, correctness, step completion time, and overall task completion status.
\end{promptblock}

\subsection{LLM Judge Prompt and Evaluation Cards}
\label{app:judge_prompts}

This appendix reports the system prompt, input format, and task-specific evaluation cards used by the LLM judge for evaluating specification propagation.

\subsubsection{Judge System Prompt}

\begin{promptblock}
You are evaluating a single multi-agent authoring session for XR study prototyping.

Your goal is to evaluate specification propagation across this chain:
Initial Prompt -> Study Design Spec -> Scene Spec -> Interaction Spec

Use only the provided task card, initial prompt, and the filtered session payload.
The filtered session payload contains:
- latest_specs.study_design_spec
- latest_specs.scene_spec
- latest_specs.interaction_spec

Evaluate specification alignment and consistency only.
Do not infer runtime execution success or code compilation success.

Use these three dimensions only:

1. Prompt-to-Study Alignment (P2D)
- P2D1: core research goal preservation
- P2D2: core task mechanic preservation
- P2D3: condition, state, or ordered-step preservation
- P2D4: core logging intent preservation

2. Study-to-Scene Grounding (D2S)
- D2S1: environment and task-space grounding
- D2S2: critical scene carrier grounding
- D2S3: condition or state carrier grounding
- D2S4: no major contradiction or drift from study design

3. Study/Scene-to-Interaction Realization (DS2I)
- DS2I1: core task flow and interaction logic realization
- DS2I2: condition, state, or step logic realization
- DS2I3: compatibility with scene spec and scene assumptions
- DS2I4: core logging realization and no major contradiction with study + scene

Score each item using:
- 0 = missing, contradictory, or clearly inconsistent
- 0.5 = partially present, underspecified, or ambiguous
- 1 = clearly present and aligned

Prioritize the task card's critical elements over secondary detail completeness.
If a core task loop and study intent are clearly preserved, do not over-penalize missing secondary UI, exact object names, or implementation detail polish.
Use semantic equivalence rather than literal naming whenever possible.
Common allowed flexibility across tasks includes script, class, and GameObject naming; exact object layout values when core spatial relations are preserved; visual presentation details; and low-level implementation detail that does not change the core task logic.
Omission of secondary elements should usually reduce a score from 1 to 0.5, not from 1 to 0, unless the omission breaks the core task or study logic.
Be strict about major drift, contradictions, and broken core traceability.

Return JSON only, following the provided schema exactly.
\end{promptblock}

\subsubsection{Judge Input Format}

\begin{promptblock}
{
  "task_name": "...",
  "initial_prompt": "...",
  "task_card": {
    "core_research_goal": "...",
    "critical_prompt_elements": ["..."],
    "critical_scene_elements": ["..."],
    "critical_interaction_logic": ["..."],
    "allowed_flexibility": ["..."],
    "major_drift_examples": ["..."],
    "scoring_note": "..."
  },
  "session_payload": {
    "latest_specs": {
      "study_design_spec": {},
      "scene_spec": {},
      "code_spec": {}
    }
  }
}
\end{promptblock}

\subsubsection{Task Card: Hand Redirection}

\begin{promptblock}
{
  "task_name": "Hand Redirection",
  "initial_prompt": "<see Appendix initial prompts>",
  "task_card": {
    "core_research_goal": "Evaluate whether users notice mismatches between real hand movement and a visually redirected virtual hand during mid-air reaching.",
    "critical_prompt_elements": [
      "mismatch detection between real and redirected virtual hand movement",
      "two-target reaching task",
      "normal versus redirected comparison",
      "post-reach consistent versus altered judgment",
      "per-trial logging of participant, condition, trial, response, accuracy, and time"
    ],
    "critical_scene_elements": [
      "two target markers or equivalent reach endpoints",
      "a user hand or redirectable virtual hand carrier",
      "a response UI or equivalent post-trial response mechanism"
    ],
    "critical_interaction_logic": [
      "trial flow for repeated reaches",
      "condition switching between normal and redirected behavior",
      "response capture after each reach",
      "response accuracy or condition-aware scoring logic",
      "per-trial logging"
    ],
    "allowed_flexibility": [
      "script, class, and GameObject naming",
      "exact object layout or placement values if core spatial relations are preserved",
      "visual presentation details",
      "low-level implementation details that do not change the core task logic"
    ],
    "major_drift_examples": [
      "replacing the task with unrelated motor task",
      "omitting the judgment step",
      "removing the two-condition comparison"
    ],
    "scoring_note": "Treat exact object naming and exact redirection implementation details as secondary if the core comparison task is preserved."
  }
}
\end{promptblock}

\subsubsection{Task Card: 3D Selection}

\begin{promptblock}
{
  "task_name": "3D Selection",
  "initial_prompt": "<see Appendix initial prompts>",
  "task_card": {
    "core_research_goal": "Build a VR prototype for 3D target pointing and selection.",
    "critical_prompt_elements": [
      "multiple 3D targets",
      "one highlighted target at a time",
      "pointing and confirmation of the highlighted target",
      "repeated trials",
      "correctness logging per trial"
    ],
    "critical_scene_elements": [
      "a set of selectable targets",
      "a visible highlight or equivalent target cue",
      "a confirmation mechanism or interaction carrier"
    ],
    "critical_interaction_logic": [
      "trial progression across multiple targets",
      "highlight target assignment",
      "selection detection",
      "correctness comparison between selected and highlighted target",
      "per-trial logging"
    ],
    "allowed_flexibility": [
      "script, class, and GameObject naming",
      "exact object layout or placement values if core spatial relations are preserved",
      "visual presentation details",
      "low-level implementation details that do not change the core task logic"
    ],
    "major_drift_examples": [
      "removing target highlighting",
      "changing into scene exploration",
      "omitting correctness tracking"
    ],
    "scoring_note": "Do not heavily penalize missing presentation polish if the highlighted-target selection loop and correctness tracking are preserved."
  }
}
\end{promptblock}

\subsubsection{Task Card: Navigation Aid / Spatial Search}

\begin{promptblock}
{
  "task_name": "Navigation Aid / Spatial Search",
  "initial_prompt": "<see Appendix initial prompts>",
  "task_card": {
    "core_research_goal": "Create an XR search task where users move through space to find a target object.",
    "critical_prompt_elements": [
      "search for a target in a city-like environment",
      "one target active in a search event",
      "no-aid versus directional-aid comparison",
      "user confirms when the target is found",
      "search-event logging with participant, condition, target, found status, time, total found, and trial index"
    ],
    "critical_scene_elements": [
      "an explorable city-like environment",
      "a target object",
      "a directional arrow or equivalent navigation aid carrier for the aided condition"
    ],
    "critical_interaction_logic": [
      "target placement or target assignment logic",
      "condition switching between no aid and aided search",
      "found confirmation logic",
      "search progression across trials or search events",
      "search-time and event logging"
    ],
    "allowed_flexibility": [
      "script, class, and GameObject naming",
      "exact object layout or placement values if core spatial relations are preserved",
      "visual presentation details",
      "low-level implementation details that do not change the core task logic"
    ],
    "major_drift_examples": [
      "removing the assisted comparison",
      "turning into static inspection",
      "omitting target confirmation logic"
    ],
    "scoring_note": "Prioritize whether the city-search, directional-aid comparison, and found-confirmation loop are preserved. Do not strongly penalize missing generic progress UI or low-level binding detail."
  }
}
\end{promptblock}

\subsubsection{Task Card: Procedural Assembly}

\begin{promptblock}
{
  "task_name": "Procedural Assembly",
  "initial_prompt": "<see Appendix initial prompts>",
  "task_card": {
    "core_research_goal": "Build an XR procedural training prototype for a simple assembly task.",
    "critical_prompt_elements": [
      "multiple parts in a workspace",
      "ordered multi-step assembly",
      "step-by-step instruction",
      "correctness check before progression",
      "per-step logging with expected action, completed action, correctness, time, and overall completion status"
    ],
    "critical_scene_elements": [
      "a workspace or assembly area",
      "multiple manipulable parts",
      "an instruction carrier",
      "a placement or assembly target area"
    ],
    "critical_interaction_logic": [
      "step sequence control",
      "part manipulation support",
      "step correctness checking",
      "progression only after valid completion",
      "per-step logging"
    ],
    "allowed_flexibility": [
      "script, class, and GameObject naming",
      "exact object layout or placement values if core spatial relations are preserved",
      "visual presentation details",
      "low-level implementation details that do not change the core task logic",
      "exact step details may vary as long as the intended step order and progression logic are preserved"
    ],
    "major_drift_examples": [
      "removing ordered progression",
      "omitting correctness checks",
      "turning the task into freeform object play"
    ],
    "scoring_note": "Focus on whether ordered procedural training is preserved. Do not over-penalize differences in assembly theme or helper mechanics if stepwise correctness logic is intact."
  }
}
\end{promptblock}

\end{document}